\documentclass[a4paper,11pt]{article}
\pdfoutput=1 % if your are submitting a pdflatex (i.e. if you have
\usepackage[T1]{fontenc} % if needed

\usepackage{cancel}
\usepackage{subcaption}
\usepackage{amsmath,amssymb}
\usepackage{graphicx}

\newcommand{\Xpr}{X^\prime_{2/3}}
\newcommand{\Xprb}{\bar{X^\prime}_{2/3}}
\newcommand{\TeV}{{\rm\ TeV}}

\begin{document}

\begin{titlepage}

\begin{center}
{\Large \bf {Interpretation of Vector-like Quark Searches: 
\\ Heavy Gluons in Composite Higgs Models}
}
\vskip 1.2cm
  {   Juan Pedro Araque$^{1}$, Nuno Filipe Castro$^{1,2}$, Jos\'{e} Santiago$^{3}$} 
\vskip 0.4cm

{\it
$^{1}$ LIP, Departamento de F\'{\i}sica, Escola de Ci\^encias, \\ Universidade do Minho, 4710-057 Braga, Portugal \\
$^{2}$ Departamento de F\'{\i}sica e Astronomia, Faculdade de Ci\^encias, \\ 
Universidade do Porto, 4169-007 Porto, Portugal\\
$^{3}$CAFPE and Departamento de F\'{\i}sica Te\'{o}rica y del
Cosmos, \\ Universidad de Granada, E-18071 Granada, Spain\\  
}

\end{center}

\vspace{0.5cm}
\begin{abstract}Pair production of new vector-like quarks in $pp$ collisions
is considered  
model independent as it is usually dominated by QCD production. We
discuss the interpretation of vector-like quark searches in the case
that QCD is not the only relevant production mechanism for the new
quarks. In particular we consider the effect of a new massive color
octet vector boson with sizeable decay branching ratio into the new
quarks. We pay special attention to the sensitivity of the Large
Hadron Collider  
experiments, both in run-1 and early run-2, to differences in the 
kinematical distributions from the different production mechanisms. We
have found that even 
though there can be significant differences in some kinematical
distributions at the parton level, the differences are washed out at
the reconstruction level. Thus, the published experimental results can be
reinterpreted in models with heavy gluons by simply rescaling the
production cross section.
\end{abstract}

\end{titlepage}

\section{Introduction}\label{sec:intro}

The discovery of the Higgs boson~\cite{Aad:2012tfa,Chatrchyan:2012xdj}
at the Large Hadron Collider (LHC)
has given us, for the first time, direct experimental access to one of the
least understood aspects of elementary particle interactions, the
mechanism of electroweak symmetry breaking (EWSB). The absence of a
dynamical mechanism that sets the scale of EWSB in the standard model
(SM) makes it difficult to conceive that the SM is the whole story. It
is far more likely that there is new physics beyond the SM reponsible
for the dynamical realization of EWSB, and a great deal of effort from
the LHC experimental collaborations is going into searching for this
possible new physics. The ingenuity of model-builders has provided us
with an enormous variety of new physics models, with a wealth of
different phenomenological signatures, to be searched for at the
LHC. The experimental collaborations have responded to this
theoretical richness in a three-fold way. First, they have made a tremendous
effort in trying to cover all possible experimental signatures of
known models. Second, they are providing more and more details of
their experimental analyses, thus making the recasting of such
analyses for different models easier. Finally, they try to interpret
their results in terms of the largest possible number of different
models so that they are as widely applicable as possible. A good
example of this are the searches for new vector-like quarks
(VLQ) that 
have gone from being searched for in only a few final states and
interpreted in terms of a single model to combining all possible final
states and reporting the corresponding bounds in terms of arbitrary
branching ratios (BR), see for
instance~\cite{Aad:2015kqa,Chatrchyan:2013uxa}. 

The last point mentioned above, the interpretation of experimental
results in terms of a limited number of theoretical models, remains
one of the main bottlenecks in the crucial exchange of information between
the theoretical and experimental communities in the search for
physics beyond the SM. Important efforts have been made recently, like
the use of simplified models~\cite{Alves:2011wf} or the
interpretation of the 
experimental results in terms of effective Lagrangians. However, even
this does not always capture the relevant information for some
well-motivated models, either because they are too complex to be
cast in the form of a reasonably simple model or because the
interpretation in terms of effective operators requires too many free
parameters or is complicated by the presence of new light particles in the
spectrum. The recent development of Monte Carlo and fast detector
simulation tools has simplified in a great manner the reinterpretation
of experimental results in terms of new models. However, a serious
analysis beyond an educated estimate requires a significant amount of
computing power and it will never reach the complexity and precision of a
full-fledged experimental analysis. 

When a new physics model has experimental signatures close to the ones
that are being searched for at the LHC it is worth studying in detail
how the corresponding analyses can be reinterpreted in terms of the
new model. There are essentially two possibilities, either the 
kinematical distributions relevant for the analyses are similar enough
(within the experimental precision) 
to the ones in one of the models that are being used for the
interpretation or some
of the distributions are significantly different. In the former case 
the reinterpretation is trivial whereas in the latter it
becomes much more complicated but on the other hand
there is a possible handle to distinguish the new model from the ones
being used for interpretation (and to increase the sensitivity of the search
to the new model). The ideal case is actually somewhere in between,
with identical distributions for the variables that have been used in
the experimental analyses but with significant differences in extra
variables that could be eventually used to design more sensitive
searches. The main goal of this article is to perform such a study in
a very well motivated model. In particular we consider the
reinterpretation of VLQ searches in terms of a new production
mechanism, namely the s-channel exchange of a massive color octet
vector boson (a heavy gluon), in composite Higgs models (CHM) with partial
compositeness. 

CHM~\cite{Kaplan:1983fs} with partial
compositeness~\cite{Kaplan:1991dc} predict new VLQ, called  
top partners, that, in the light of the Higgs mass, are expected to be
relatively light~\cite{Matsedonskyi:2012ym}, with masses in the TeV region. 
The mechanism of partial compositeness, that
implies a linear mixing between the elementary and composite sectors,
guarantees the presence of colored resonances in the spectrum and in
particular, massive color octet vectors are naturally expected. If
these heavy gluons have similar masses to their electroweak
counterparts, they have to be in the multi-TeV range (due to stringent
constraints from electroweak precision data on the
latter~\cite{Agashe:2003zs}) and 
therefore it is likely that their decay into top partners, which is
favored by large couplings, is kinematically open. In this case the
heavy gluon will have a sizeable width and will decay mainly into the
top partners, in pairs if kinematically allowed, thus significantly
reducing the sensitivity of $t\bar{t}$ resonance searches to the heavy
gluon~\cite{Chala:2014mma}\footnote{The case in
which the heavy gluon decays into a SM quark and a VLQ has been
studied in~\cite{Barcelo:2011wu,Bini:2011zb}.}. 
In this case VLQ searches can be the most
sensitive probes of the heavy gluon (see~\cite{Azatov:2015xqa} for a
recent study in which bounds on the heavy gluon have been computed by
recasting experimental searches for pair production of new charge
$5/3$ VLQ and~\cite{Vignaroli:2015ama} for implications on recent
reported excesses at the LHC.). This is therefore an
example in which we have the same final state considered in several
experimental searches (VLQ pair production) but with a new production
mechanism. The goal of our study is to investigate if the large mass of the
heavy gluon has a significant impact on the kinematical distributions
used in VLQ seaches. We will see that the relevant distributions, which
depending on the region of parameter space can be significantly
different at the
parton level, are very similar to the ones in which only QCD
production is considered, once detector and reconstruction effects are
taken into account. Thus, in most of the parameter space, VLQ searches
can be trivially reinterpreted in terms of a rescaling of the cross
section in models with heavy gluons
without the need of costly computer simulations. One can simply use
the published experimental limits on the VLQ production cross section
as a function of the VLQ mass, that assume QCD production, and
overlay the theoretical production cross section in the model with
heavy gluons to compute the limits. We then use this approach to
compute the current bounds on models with a heavy gluon as derived
from the LHC run-1 VLQ searches and estimate the sensitivity of the early
LHC run-2 data.

The rest of the article is organized as follows. We introduce the
model in Section~\ref{model}. We compare the relevant kinematical
distributions in a model in which QCD is the only production mechanism
with the ones in our model in Sections~\ref{parton} 
at truth parton level and~\ref{reconstruction} 
including detector and reconstruction effects. We then
discuss the effect that the differences between the kinematical
distributions in both models have on the actual limits extracted in
Section~\ref{limits:two:ways}. 
Section~\ref{limits} is dedicated to present the current
limits on the model and early run-2 expectations and we conclude in
Section~\ref{conclusions}. Technical details of the model are
presented in Appendix~\ref{apend:model}. 

\section{The Model \label{model}}

CHM with partial compositeness represent a well-motivated class of
models of dynamical EWSB. They usually predict a number of relatively
light VLQ with large couplings to the top (and bottom) quark,
the so-called top partners, and also new vector resonances, possibly
in the octet representation of the color group that couple sizeably to
the top quark and its partners. They are therefore a prime candidate
to study the interplay of VLQ and heavy gluon searches that
we mentioned in the introduction. For the sake of concreteness we have
considered a simplified version of the 
Minimal CHM~\cite{Agashe:2004rs}, based on the
$SO(5)/SO(4)$ coset, with a fully composite
right-handed (RH) top quark~\cite{DeSimone:2012fs}.
In particular we take the MCH4$_5$ model in which the SM left-handed
(LH) quarks are embedded in a $5$ of $SO(5)$ and the top partners span
a $4$ representation of $SO(4)$. Specifically they read, in the basis
we are considering,
\begin{equation}
(Q^5_L)_I = \frac{1}{\sqrt{2}} 
\begin{pmatrix} 
\mathrm{i} b_L \\
 b_L \\
\mathrm{i} t_L \\
-t_L \\
0
\end{pmatrix},
\qquad
\Psi^i = \frac{1}{\sqrt{2}}
\begin{pmatrix}
\mathrm{i}(B-X_{5/3}) \\
B+X_{5/3} \\
\mathrm{i}(T+X_{2/3}) \\
-T+X_{2/3}
\end{pmatrix},
\end{equation}
where $I=1,\ldots,5$ and $i=1,\ldots,4$, respectively. 
In terms of $SU(2)_L\times U(1)_Y$ representations, the $4$ of $SO(4)$
$\Psi$ gives rise to two doublets with hypercharges $1/6$, $(T,B)$,
and $7/6$, $(X_{5/3}, X_{2/3})$, respectively. The latter contains an exotic state
with charge 5/3, $X_{5/3}$, and a charge 2/3 state, $X_{2/3}$ whereas
$T$ and $B$ have electric charge $2/3$ and $-1/3$, respectively.
After EWSB, a linear combination of $T$ and $X_{2/3}$, that we denote 
$X^\prime_{2/3}$ remains degenerate with $X_{5/3}$. The orthogonal
combination, that we call $T^\prime$, and $B$ are somewhat heavier with
a small mass splitting. In most of the parameter space their decay
BR read
\begin{eqnarray}
BR(X_{5/3}\to t W^+)
&=&
BR(B\to t W^-)=1, \\
BR(X^\prime_{2/3}\to t Z)
&\approx&
BR(X^\prime_{2/3}\to t H)
\approx
\frac{1}{2}, \\
BR(T^\prime \to t Z)
&\approx&
BR(T^\prime \to t H)
\approx \frac{1}{2}.
\end{eqnarray}

The new ingredient with respect to the MCH4$_5$ model 
is the presence of the heavy gluon. We
introduce it following the partial compositeness
mechanism~\cite{Chala:2014mma}.
The relevant part of the Lagrangian, in the
elementary-composite basis, reads 
\begin{eqnarray}
\mathcal{L} &=& 
\bar{q}_L\mathrm{i}\cancel{D} q_L 
+\bar{t}_R\mathrm{i}\cancel{D} t_R
+\bar{\Psi}\mathrm{i}(\cancel{D}+\mathrm{i} \cancel{e}) \Psi
- M_\Psi \bar{\Psi} \Psi 
\nonumber \\
&+&
\Big[ \mathrm{i} c_1 (\bar{\Psi}_R)_i \gamma^\mu d^i_\mu t_R
+ y f (\bar{Q}^5_L)^I U_{Ii} \Psi^i_R 
+ y c_2 f (\bar{Q}^5_L)^I U_{I5} t_R 
+\mathrm{h.c.}
\Big]
\nonumber \\
&-&
\frac{1}{2} \mathrm{Tr}[G^e_{\mu \nu}]^2
-\frac{1}{2} \mathrm{Tr}[G^c_{\mu \nu}]^2
+\frac{1}{2} M_c^2 \left( G^c_\mu -\frac{g_e}{g_c} G^e_\mu\right)^2.
\label{Lag:MCH45}
\end{eqnarray}
This Lagrangian is determined by three dimensionless (order one)
couplings, $y$ and $c_{1,2}$, the strong couling scale $f$, the
composite gluon mass $M_c$ and the ratio of elementary to composite
gluon couplings $g_e/g_c$. The explicit expressions of
the different operators in Eq. (\ref{Lag:MCH45}) are given in
Appendix~\ref{apend:model}. 
The relevant feature for us is that the elementary gluon
only couples to elementary quarks, with coupling $g_e$, and the
composite gluon only couples to the composite quarks with coupling
$g_c$. (For simplicity we have assumed a universal
  coupling of the composite gluon 
  to the composite states. This assumption does not
  have any relevant implication on the main conclusions of our study.)
The elementary-composite gluon system can be brought to the physical
basis by means of the following rotation
\begin{equation}
\begin{pmatrix}
G^e_\mu \\ G^c_\mu 
\end{pmatrix}
=
\begin{pmatrix}
\cos \theta_3 & -\sin \theta_3 \\
\sin \theta_3 & \cos \theta_3
\end{pmatrix}
\begin{pmatrix}
g_\mu \\ G_\mu
\end{pmatrix},
\end{equation}
where the ratio of couplings fixes the mixing angle
$\tan\theta_3=\frac{g_e}{g_c}$.
After this rotation we have a 
massless color octet, the SM gluon $g_\mu$, 
and a heavy gluon, $G_\mu$, with mass 
\begin{equation}
M_G=\frac{M_c}{\cos \theta_3}.
\end{equation}
The SM gluon couples universally with coupling strength
\begin{equation}
g_s=g_c \sin\theta_3=g_e \cos\theta_3,
\end{equation}
whereas the heavy gluon has couplings to elementary and composite fields
given by
\begin{equation}
G \bar{\psi}_{\mathrm{elem}} \psi_{\mathrm{elem}} :
-\frac{g_s^2}{\sqrt{g_c^2 - g_s^2}}, \qquad
G \bar{\psi}_{\mathrm{comp}} \psi_{\mathrm{comp}} :
\sqrt{g_c^2 - g_s^2}.
\end{equation}

In the following we fix the input parameters to the following
values:
\begin{equation}
g_c=3, \qquad f=800\mbox{ GeV}, \qquad c_1=0.7, \qquad c_2=1.7~,
\end{equation}
where the value of $g_c$ is fixed only to test the independence of the
relevant kinematical distributions of the reconstructed objects on the
presence of the heavy gluon, described in detail in the next
section. In order to get the bounds on $M_G$ from current data and the
expected reach with the early LHC run-2 data we will vary $2\leq g_c \leq 5$. 
$y$ is fixed by the top mass and its value determines the degree of
compositeness of the LH top. It is most sensitive to the value of
$c_2$ and has a milder dependence on the mass of the top parners.
For the values of the input parameters chosen above it ranges from
$y\approx 0.95$ for $M_\Psi=600$ GeV to $y \approx 0.61$ for $M_\Psi =
1.6$~TeV. This in turns corresponds to a degree of compositeness for
the LH top quark 
\begin{equation}
s_L= \frac{y f}{\sqrt{y^2 f^2+M_\Psi^2}},
\end{equation}
of $s_L=0.78$ and $s_L=0.29$, respectively. As it will become clear in
the following sections, our goal is to show that the kinematical
distributions of the reconstructed objects in VLQ searches are not
affected in a significant way by the presence of the heavy gluon. This
effect is not very sensitive to the degree of compositeness of the LH
top quark, a fact that we implicitly check by not keeping $s_L$ fixed
in our analyses.

The only remaining parameters are the values of the top partners and
heavy gluon masses, $M_\Psi$ and $M_G$, and the behaviour of the light
generations. Regarding the masses we will vary them in the relevant
range 
\begin{equation}
600\mbox{ GeV} \leq M_\Psi \leq 1.6\mbox{ TeV},
\qquad
1.5\mbox{ TeV} \leq M_G \leq 4.5\mbox{ TeV}.
\end{equation}
Finally, for simplicity we assume that the first two generations and
the RH bottom quark are purely elementary.

One final comment to fully characterize our model is the fact that,
due to the large number of decay channels into top partners and the
large coupling to them, the width of the heavy gluon is typically very
large, becoming quite easily comparable to $M_G$ itself. In that
circumstance, it is compulsory to consistently include the full
energy-dependent quantum
width~\cite{Barcelo:2011wu,Barcelo:2011vk,Azatov:2015xqa}. Due to the rapidly
decaying parton distribution functions (PDF), a larger width for the
heavy gluon makes it more similar to the continuum QCD production of
the top partners than a narrower one. Thus, since we are trying to
find how different the VLQ pair production via the heavy gluon is from the QCD
one, we have considered a modification of the model in which the
couplings to the composite fermions, including the RH top and the LH
top and bottom, are rescaled down, when necessary, 
to fix the maximum width of the
heavy gluon to $20\%$ of its mass. We emphasize again that this
variation goes in the direction of making the eventual differences
between the two production mechanisms larger and therefore if no
significant differences are found in this case, they are not expected
to arise in the usual case with a wider heavy gluon.~\footnote{With
this modification, the heavy gluon has a behaviour more similar to the
narrower electroweak vector boson resonances, recently studied
in~\cite{Greco:2014aza}. These resonances might be responsible for the 
reported diboson excess if the decay into top partners is not
open~\cite{Carmona:2015xaa}.}

\section{Comparison of kinematical distributions and reinterpretation of
limits}

In this section we want to investigate whether the presence of a heavy
gluon in CHM affects VLQ pair production in a significant way. If it
does, detailed simulations are needed to recast the result of current VLQ
searches for the model at hand, although then new experimental searches
might be designed to gain further sensitivity to these models. 
If it does not, the bounds can be
trivially reinterpreted in CHM with heavy gluons by simply rescaling the
corresponding cross section. 
In the case of a heavy gluon, one could expect that its large mass
provides the VLQ with a larger energy and therefore modify the
kinematical distributions of the particles they decay into. Clearly, how
large the effect is depends on the relation between the heavy gluon
and the VLQ masses and also on the available energy. 

In order to test all these features in detail we will concentrate on
one particular analysis, the VLQ pair-production searches in the $Zt$
channel~\cite{Aad:2014efa} and
display the kinematical distributions
assuming production only of $X^\prime_{2/3}$. In Section~\ref{parton}
we will show the differences on the main kinematical distributions at
the parton level. For a better comparison we will display the
distributions both normalized to their corresponding 
cross section and also to unit area. 
This latter normalization gives us a more clear picture of
whether the differences can be interpreted as purely cross section
rescaling or they present significant shape differences. 
We will then compare the distributions after hadronization, showering
and detector simulation is included in
Section~\ref{reconstruction}. Finally, in
Section~\ref{limits:two:ways} we compare the current limits from run-1
data with a full recasting of the analysis in the model versus a
simple rescaling of the corresponding cross sections.

We have generated a UFO model~\cite{Degrande:2011ua} 
using \texttt{FeynRules}~\cite{Alloul:2013bka}. The
simulations at parton level have been done
with \texttt{MG5}~\cite{Alwall:2014hca} while
for hadronization/showering and detector simulation we have
used \texttt{pythia 6}~\cite{Sjostrand:2006za} and \texttt{Delphes
3}~\cite{deFavereau:2013fsa}, 
respectively.
We have used the default
ATLAS \texttt{Delphes} card with the following modified parameters:
the \texttt{FastJet}~\cite{Cacciari:2011ma} 
ParameterR and the $\Delta R$ in the 
$b$-jets have been set to $0.4$; the $b$-tagging efficiency has been
set to $70\%$ and the mistag probabilities for $c$-jets and
light-quark/gluon jets to $20\%$ and $0.7\%$, respectively.

\subsection{Kinematical differences at parton level\label{parton}}

In this section we explore the kinematical differences, at the parton
level, 
between the
production of a pair of vector-like quarks via QCD and considering the
presence of a heavy gluon in the s-channel.  
In order to disentangle the origin of any possible kinematical
differences, we compare the pair production of $X^\prime_{2/3}$  
quarks via QCD (QCD), via the heavy gluon only (HG) and via both QCD
and heavy gluon, including interference effects (QCD+HG). 
We have studied a large number of kinematical distributions
for different values of the $G$ and $X^\prime_{2/3}$ masses, both at
$\sqrt{s}=8$ and $13$ TeV and we will report on the most relevant ones. 
When showing kinematical distributions in this section, we will show the
distributions normalized to their corresponding cross section on the
left and the same distributions normalized to unit area on the
right. The former give us an idea of the size of the different
contributions whereas the latter give us intuition on how the
efficiencies might change depending on the production mechanism. 

\begin{figure}[ht!]
\begin{center}
\begin{subfigure}[b]{0.4\textwidth}
\includegraphics[width=\textwidth]{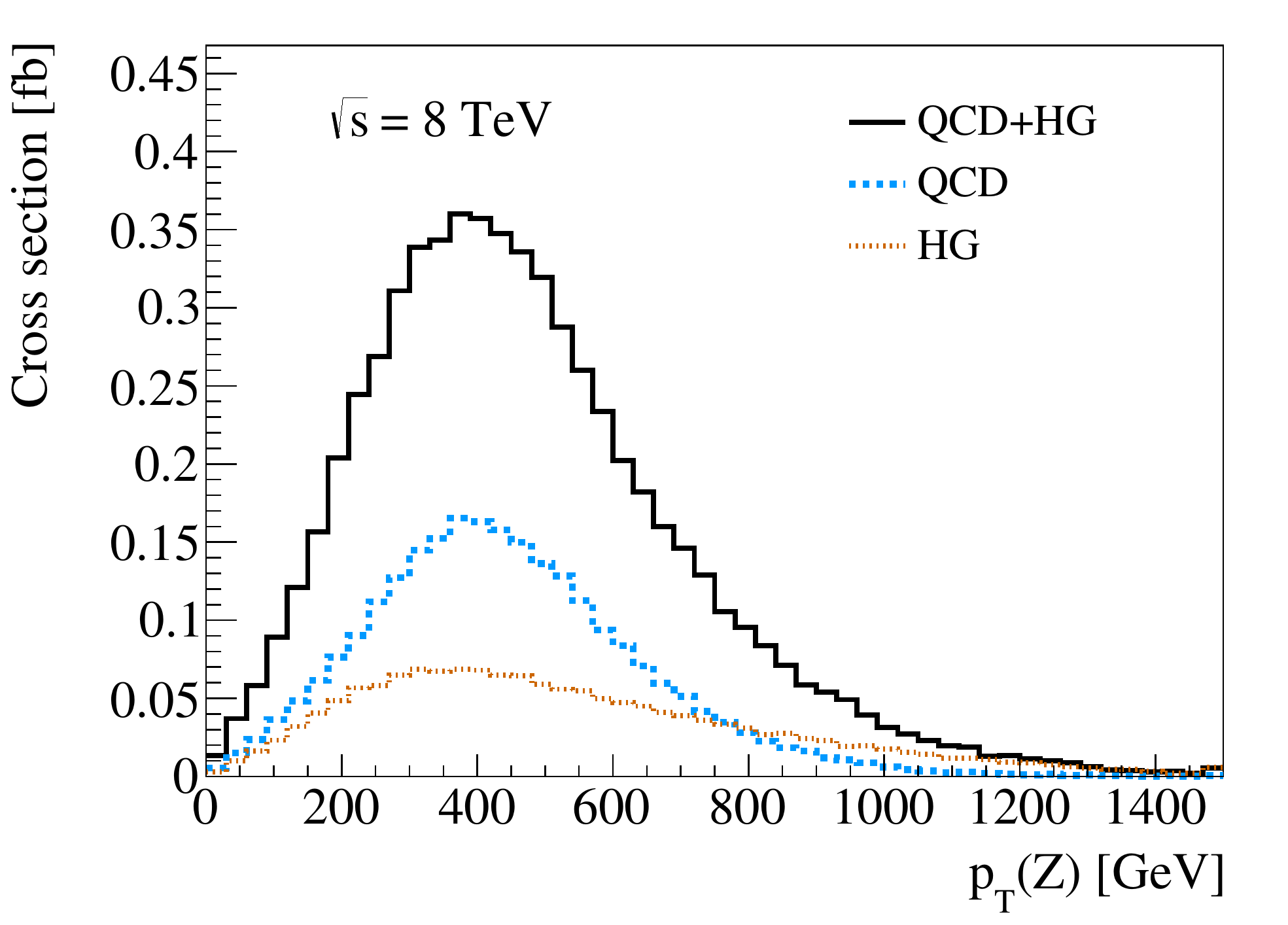}
\caption{}
\end{subfigure}
\begin{subfigure}[b]{0.4\textwidth}
\includegraphics[width=\textwidth]{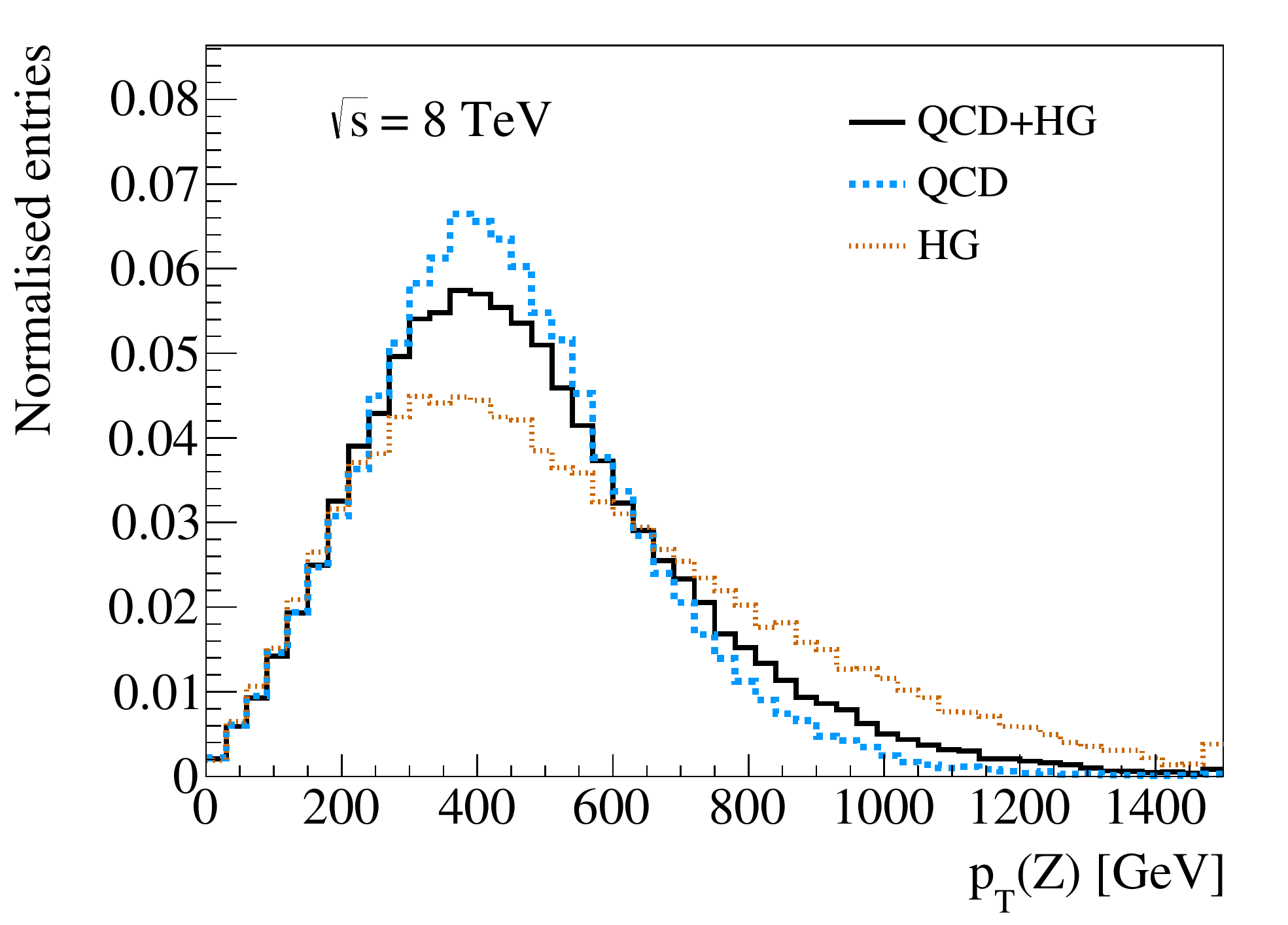}
\caption{}
\end{subfigure}\\
\begin{subfigure}[b]{0.4\textwidth}
\includegraphics[width=\textwidth]{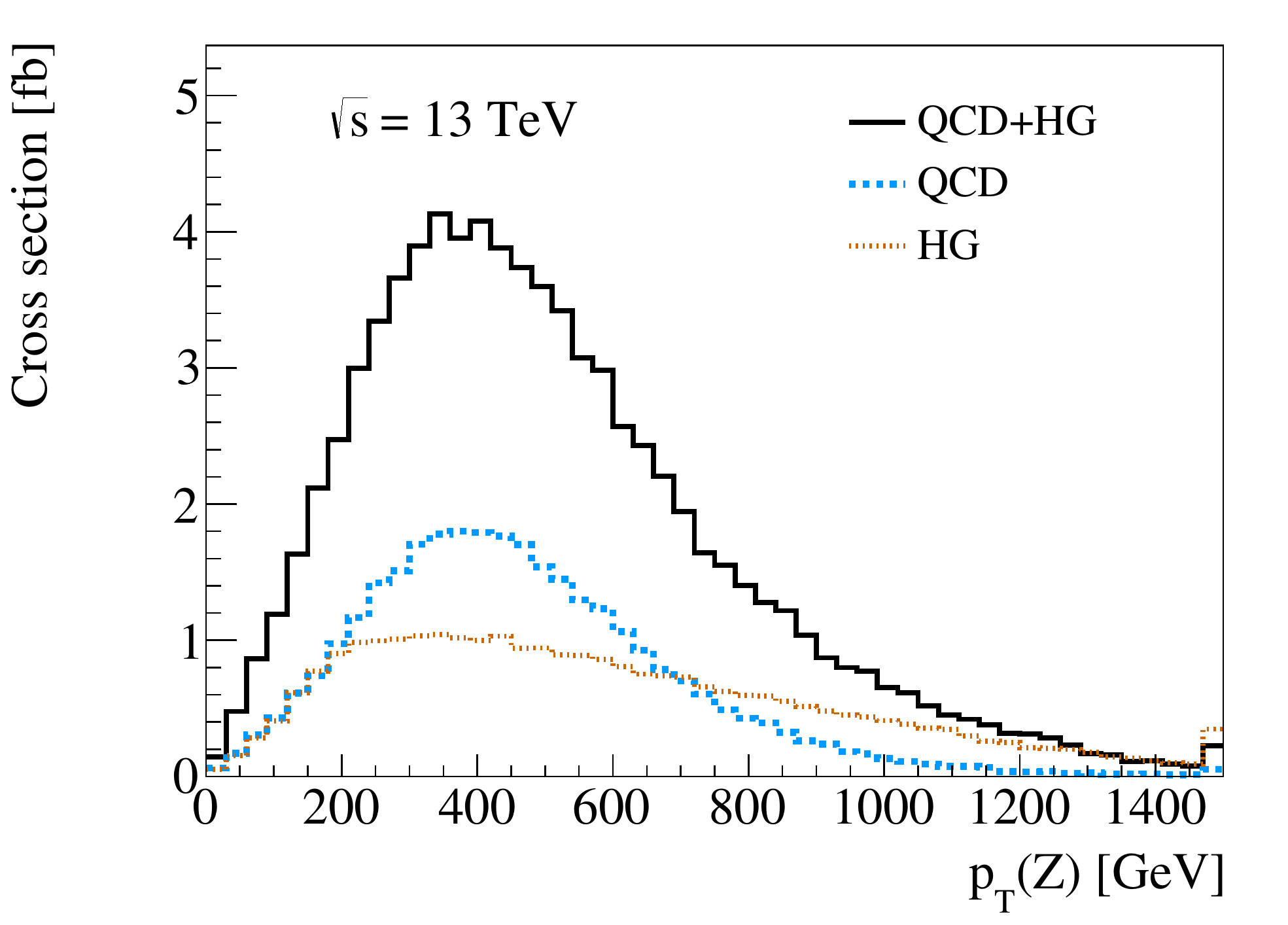}
\caption{}
\end{subfigure}
\begin{subfigure}[b]{0.4\textwidth}
\includegraphics[width=\textwidth]{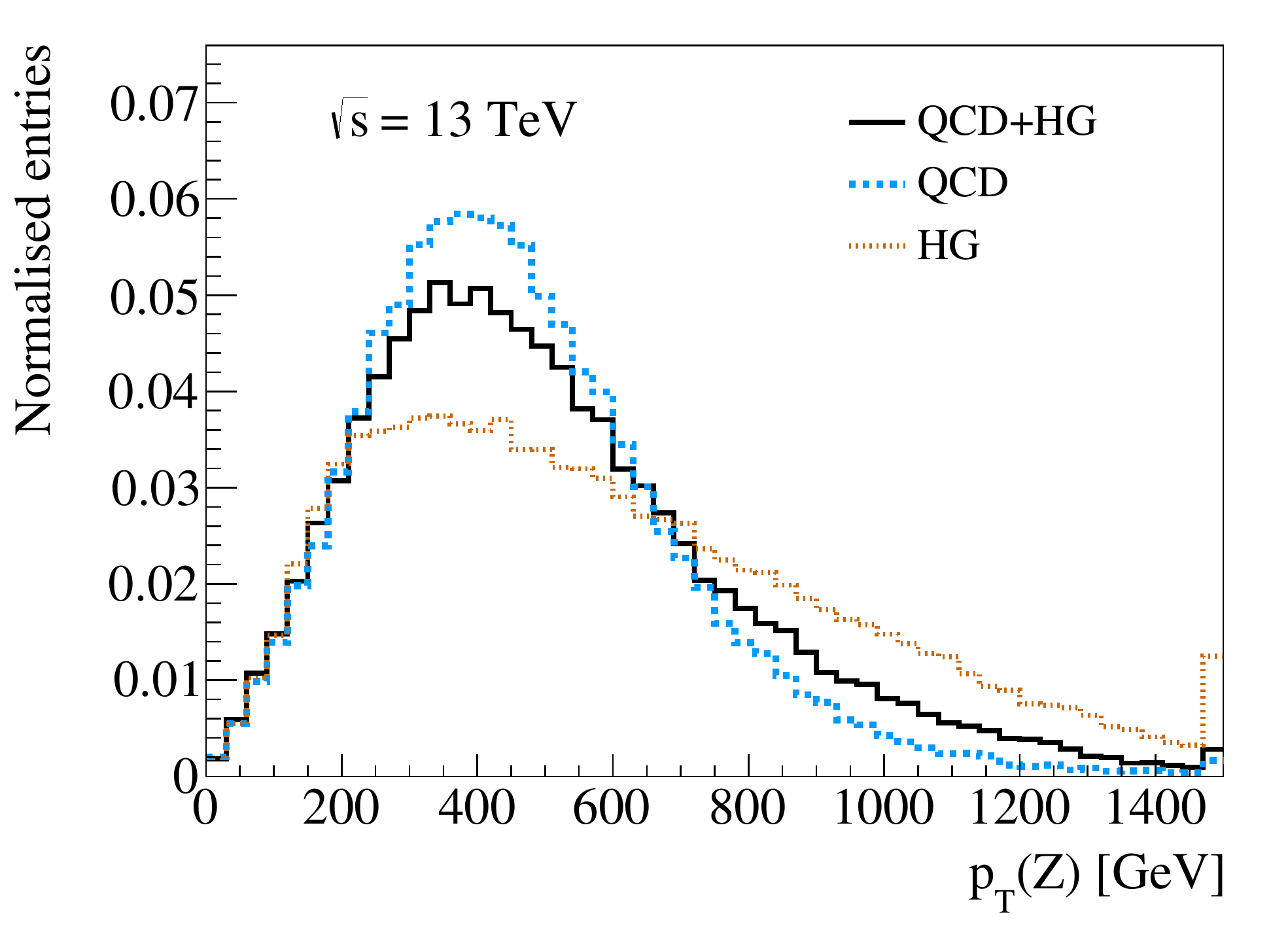}
\caption{}
\end{subfigure}
\end{center}
\caption{Distribution of the $Z$ boson transverse momentum
for $\sqrt{s}=8$ TeV (top) and $13$ TeV (bottom) at parton level with
$M_G=3.5$ TeV and $M_{\Xpr}=1$ TeV. 
The plots on the left are
normalized to the corresponding cross sections whereas the ones on the
right are normalized to unit area.}
\label{fig:Zpt}
\end{figure}

\begin{figure}[ht!]
\begin{center}
\begin{subfigure}[b]{0.4\textwidth}
\includegraphics[width=\textwidth]{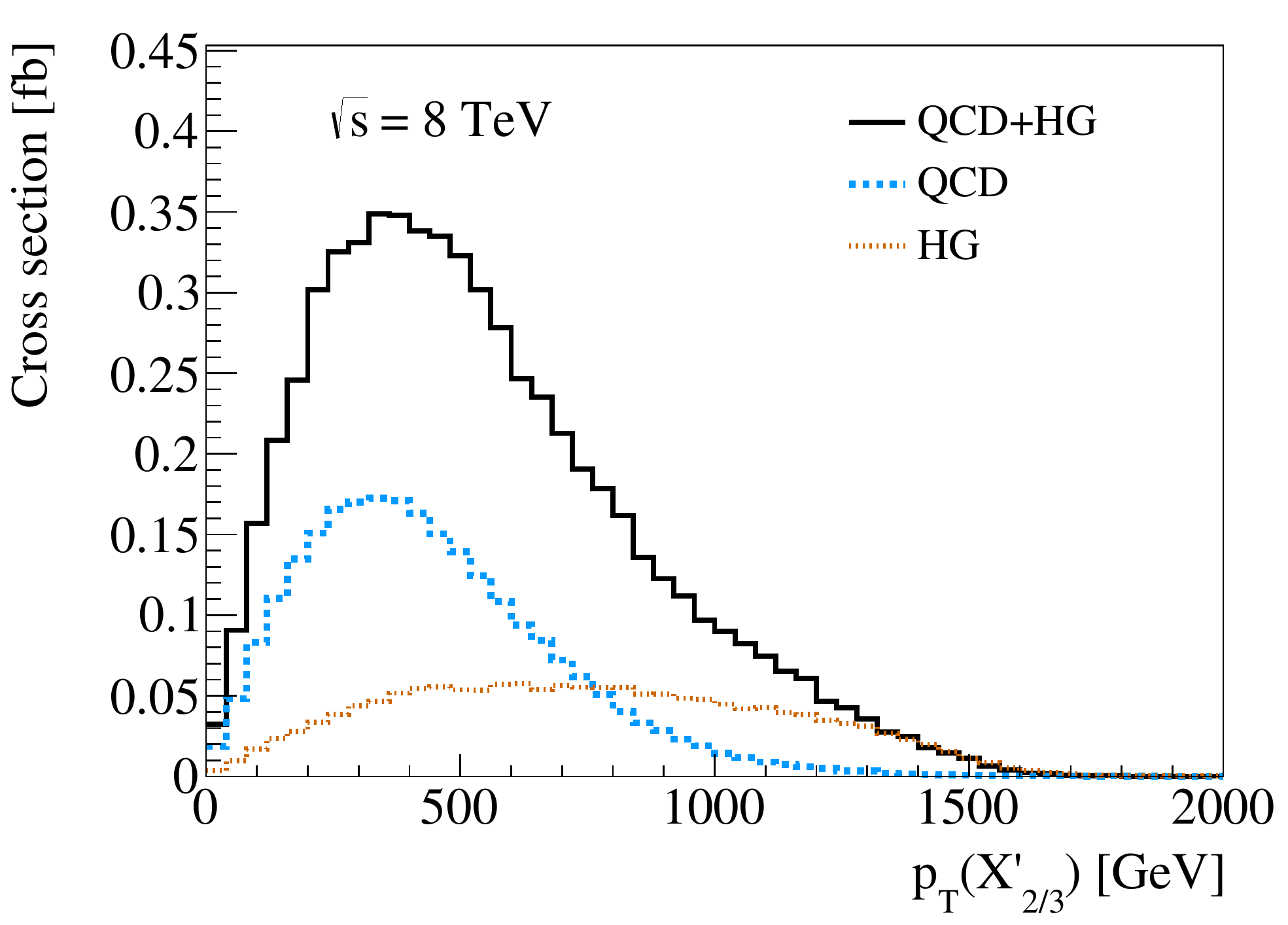}
\caption{}
\end{subfigure}
\begin{subfigure}[b]{0.4\textwidth}
\includegraphics[width=\textwidth]{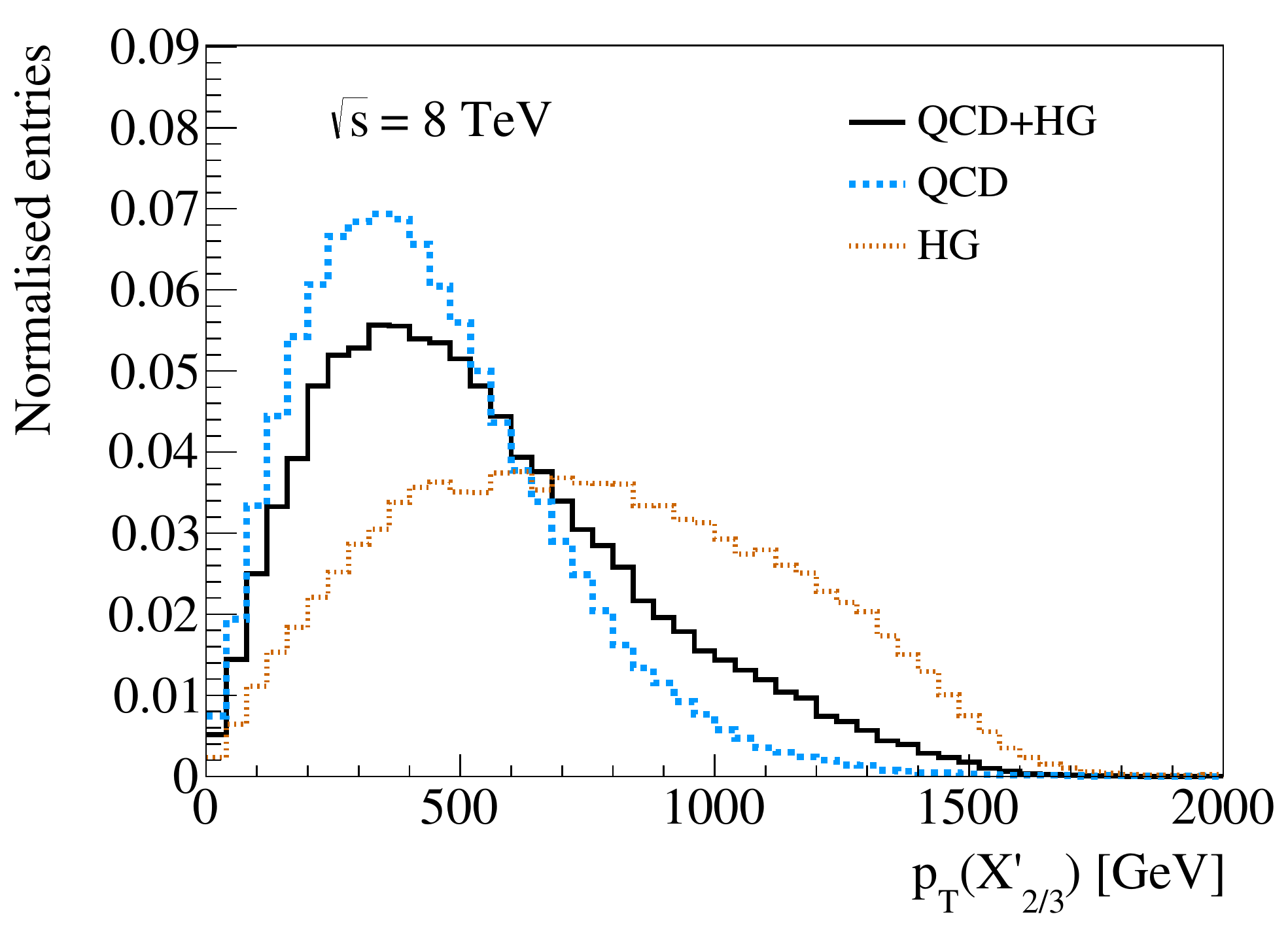}
\caption{}
\end{subfigure}\\
\begin{subfigure}[b]{0.4\textwidth}
\includegraphics[width=\textwidth]{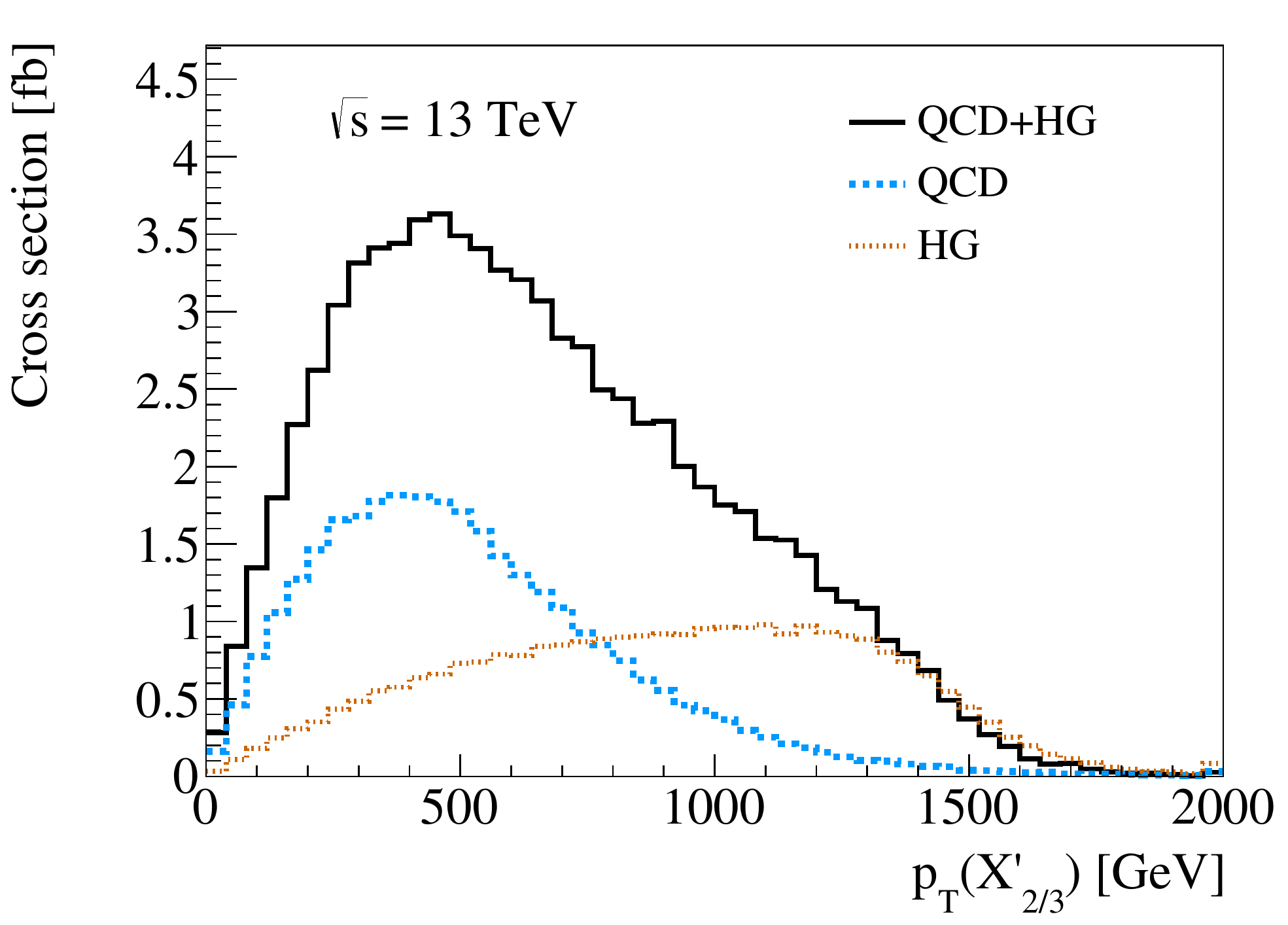}
\caption{}
\end{subfigure}
\begin{subfigure}[b]{0.4\textwidth}
\includegraphics[width=\textwidth]{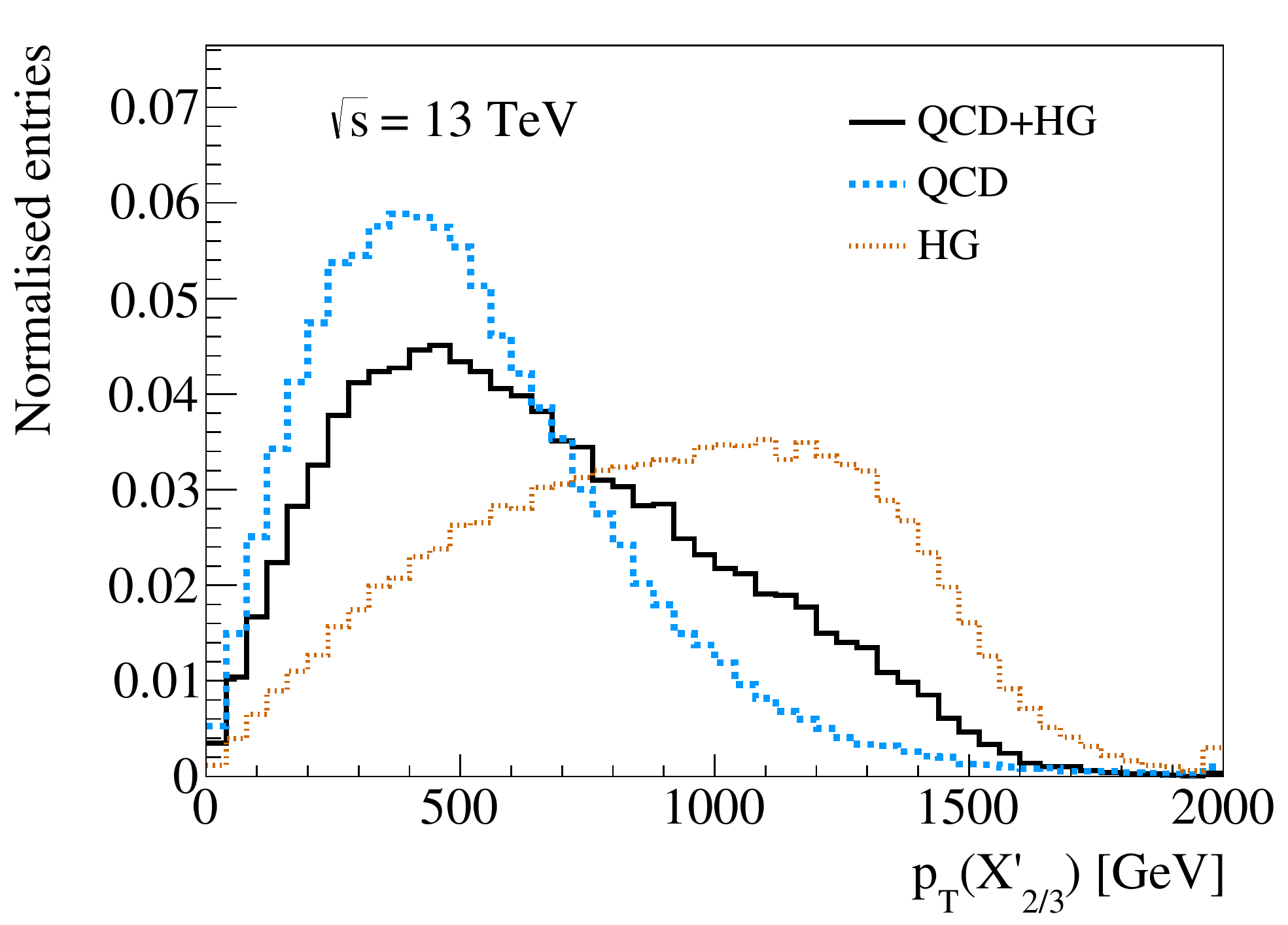}
\caption{}
\end{subfigure}
\end{center}
\caption{$\Xpr$ $p_{\mathrm{T}}$ distribution for $\sqrt{s}=8$ (top) and
$13$ TeV (bottom) at parton level with
$M_G=3.5$ TeV and $M_{\Xpr}=1$ TeV. 
The plots on the left column are normalized to the
corresponding cross section whereas the ones on the right are
normalized to unit area.}
\label{fig:Xpt}
\end{figure}

\begin{figure}[ht!]
\begin{center}
\begin{subfigure}[b]{0.4\textwidth}
\includegraphics[width=\textwidth]{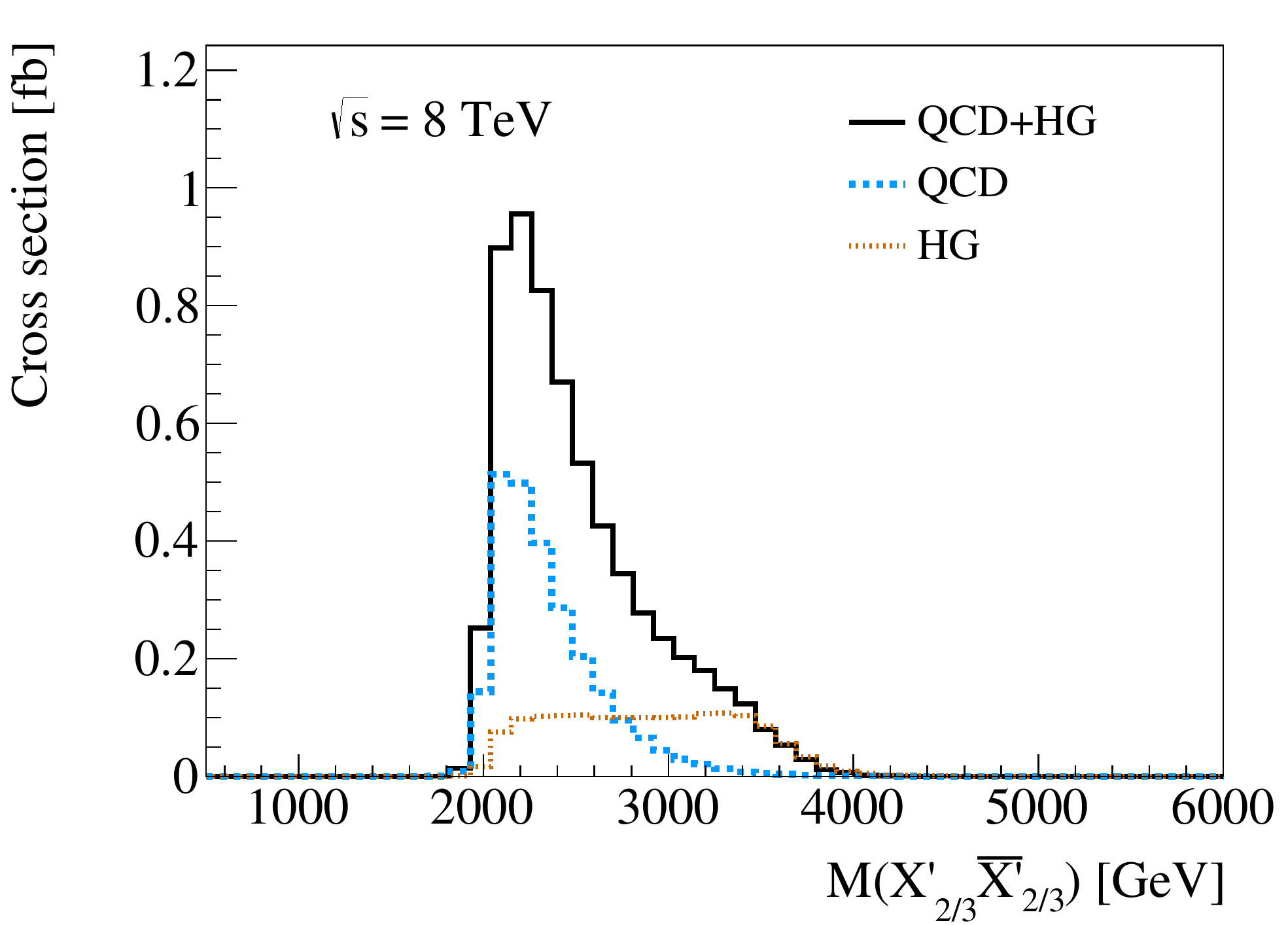}
\caption{}
\end{subfigure}
\begin{subfigure}[b]{0.4\textwidth}
\includegraphics[width=\textwidth]{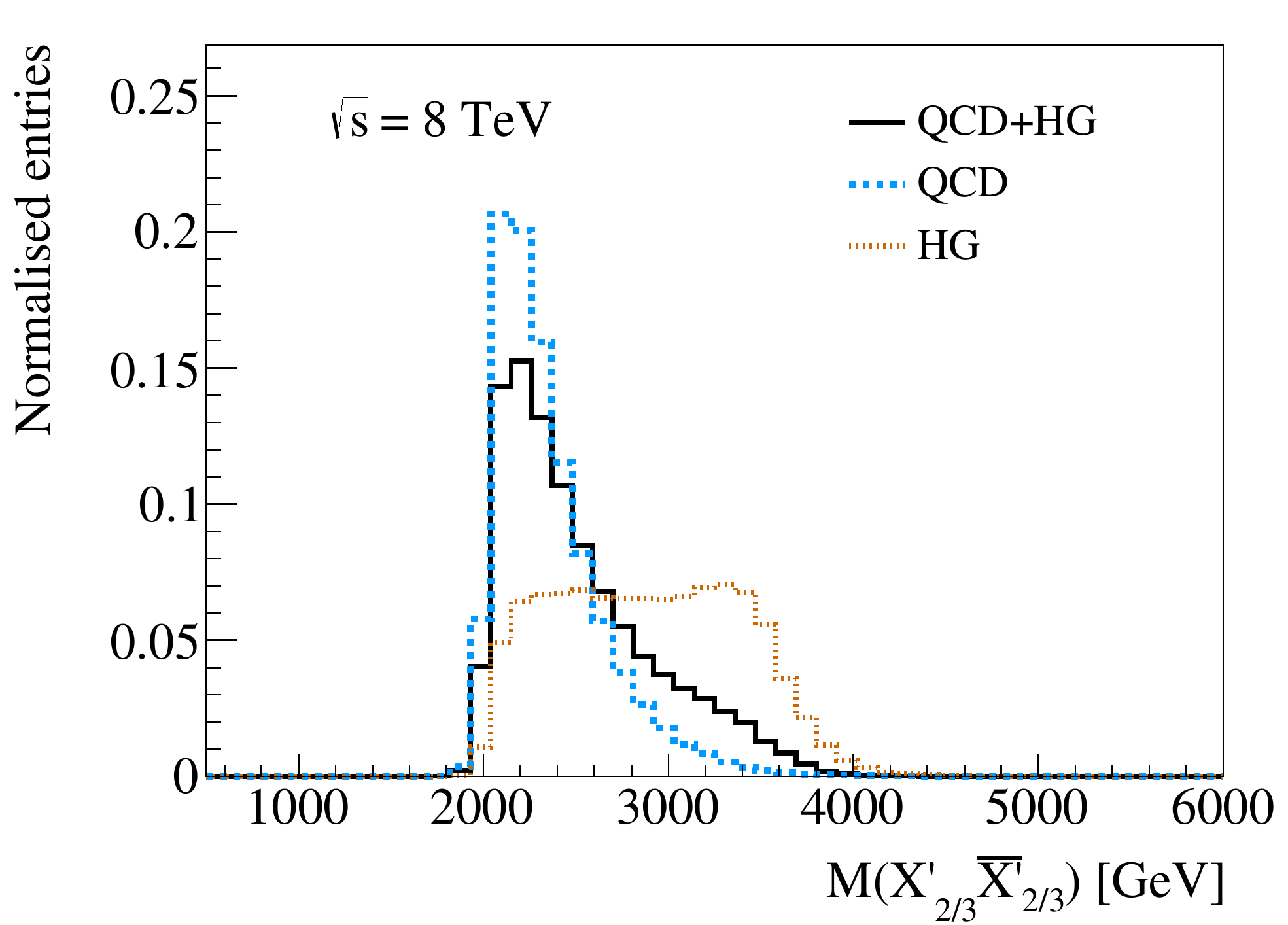}
\caption{}
\end{subfigure}\\
\begin{subfigure}[b]{0.4\textwidth}
\includegraphics[width=\textwidth]{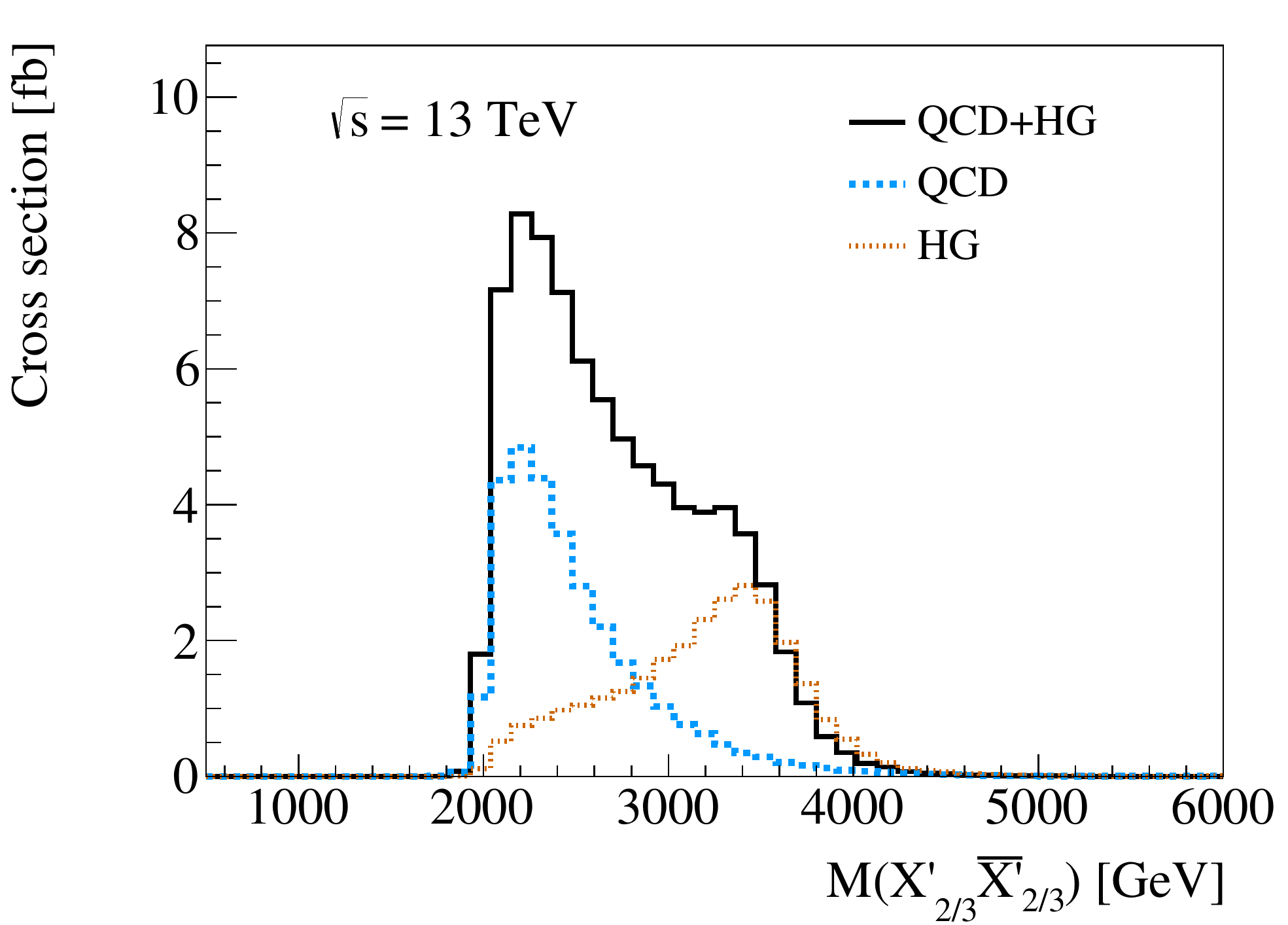}
\caption{}
\end{subfigure}
\begin{subfigure}[b]{0.4\textwidth}
\includegraphics[width=\textwidth]{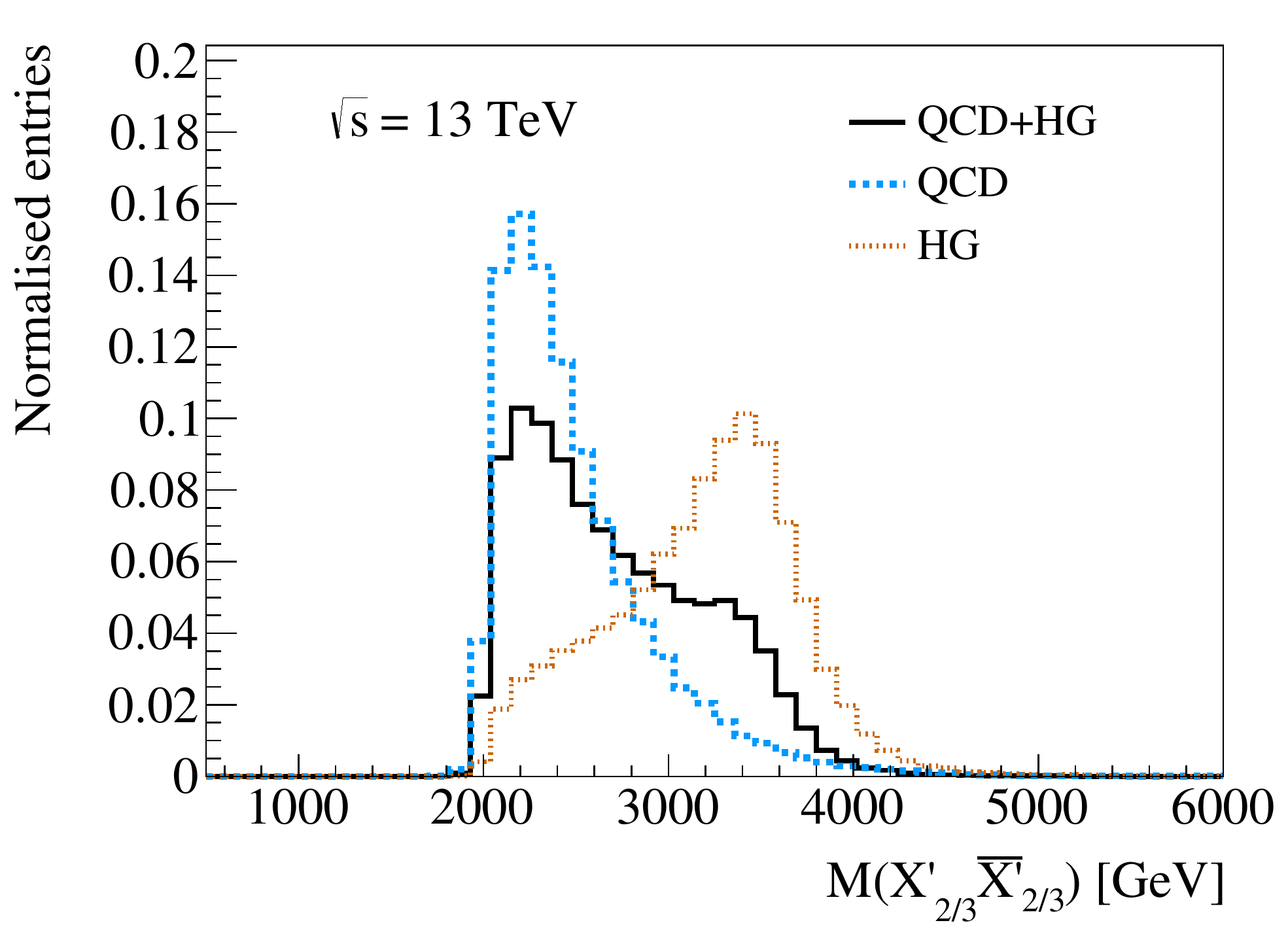}
\caption{}
\end{subfigure}
\end{center}
\caption{$\Xpr \Xprb$ invariant mass distribution for
$\sqrt{s}=8$ (top) and 
$13$ TeV (bottom)  at parton level with
$M_G=3.5$ TeV and $M_{\Xpr}=1$ TeV.  
The plots on the left column are normalized to the
corresponding cross section whereas the ones on the right are
normalized to unit area.}
\label{fig:MXX}
\end{figure}

In experimental searches it is common to start looking at
distributions of objects in the final state that are easy to
reconstruct. An example would be the $p_{\rm T}$ of the $Z$ boson 
in the $Z$ decay of the $\Xpr$ 
($\Xpr\rightarrow Zt\rightarrow ZWb$). 
We show the $p_{\rm T}(Z)$ distribution in Fig.~\ref{fig:Zpt} for 
$\sqrt{s}=8$  and 13 TeV at the top and bottom rows, respectively. 
As mentioned above, the left plots
are normalized to the corresponding cross section whereas the right
ones are normalized to unit area. We have fixed 
$M_{G}=3.5$~TeV and $M_{\Xpr} = 1$~TeV
to have a sizeable contribution from the HG channel. It 
can be seen that, even though there are clear differences between QCD
and HG, the \textit{shapes} of QCD and QCD+HG are very
similar, both at $8$ and $13$~TeV. 
We have observed a similar behaviour in properties of objects near the
detector such as the $p_{\rm T}$ of leptons or other vector bosons present
in the decay chain.   

As one could expect, kinematical distributions of objects further up the decay
chain are more sensitive to the presence of the heavy gluon.
We show in Figs.~\ref{fig:Xpt} and~\ref{fig:MXX} the distribution of 
$p_{\rm T}(\Xpr)$ and $M(\Xpr \bar{\Xpr})$, respectively. The same
layout is used with data at 8 (13) TeV in the top (bottom) row. 
The left panels normalized to the corresponding cross sections
whereas the right panels are normalized to unit area and the same
values of the $G$ and $\Xpr$ masses are chosen. As we see in the
figures, the transverse momentum of the VLQ already shows some
differences in shape between QCD and QCD+HG, differences that, as
expected, are even larger for the invariant mass of the VLQ pair.
The differences are also larger at 13~TeV than at 8~TeV as can be
easily understood from Fig.~\ref{fig:MXX}.
Due to the PDF suppression, a heavy gluon with $M_G=3.5$~TeV 
is largely produced off-shell for $\sqrt{s}=8$~TeV. 
At 13~TeV, on the other hand, the heavy gluon is mostly
produced on-shell, with a distinctive peak
in the invariant mass distribution. The kinematical distributions of its
decay products are therefore harder and easier to distinguish from the
QCD case. Of course, there is a balance between distributions that are
more easily measured but less sensitive and those that are more
sensitive to the presence of the heavy gluon but more difficult 
to measure. Furthermore, the latter will naturally suffer a larger
degradation when more realistic effects, like hadronization and
detector simulation are taken into account, as discussed in
Section~\ref{reconstruction}. 

\begin{figure}[t]
\begin{center}
\begin{subfigure}[b]{0.4\textwidth}
\includegraphics[width=\textwidth]{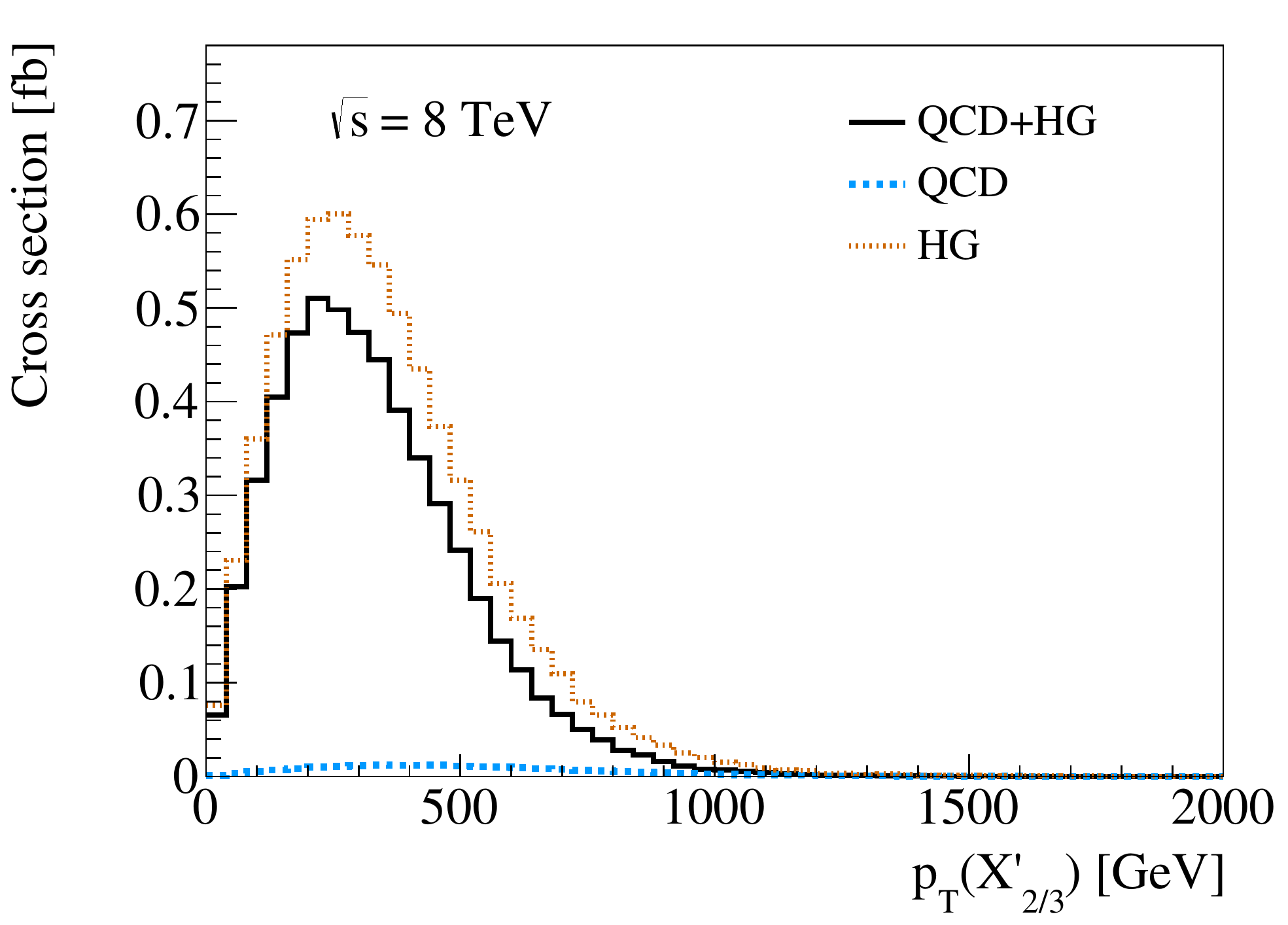}
\caption{}
\end{subfigure}
\begin{subfigure}[b]{0.4\textwidth}
\includegraphics[width=\textwidth]{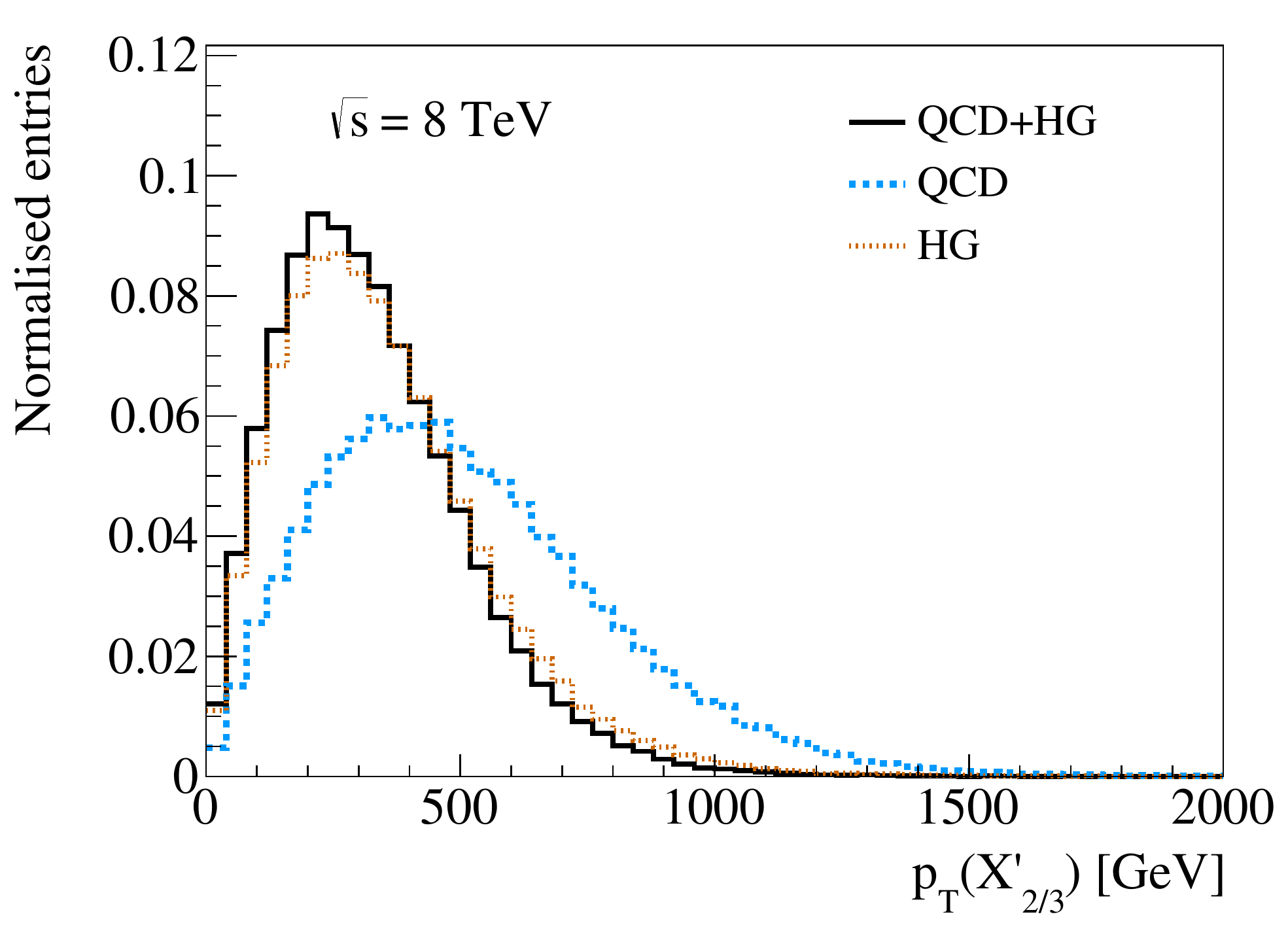}
\caption{}
\end{subfigure}\\
\begin{subfigure}[b]{0.4\textwidth}
\includegraphics[width=\textwidth]{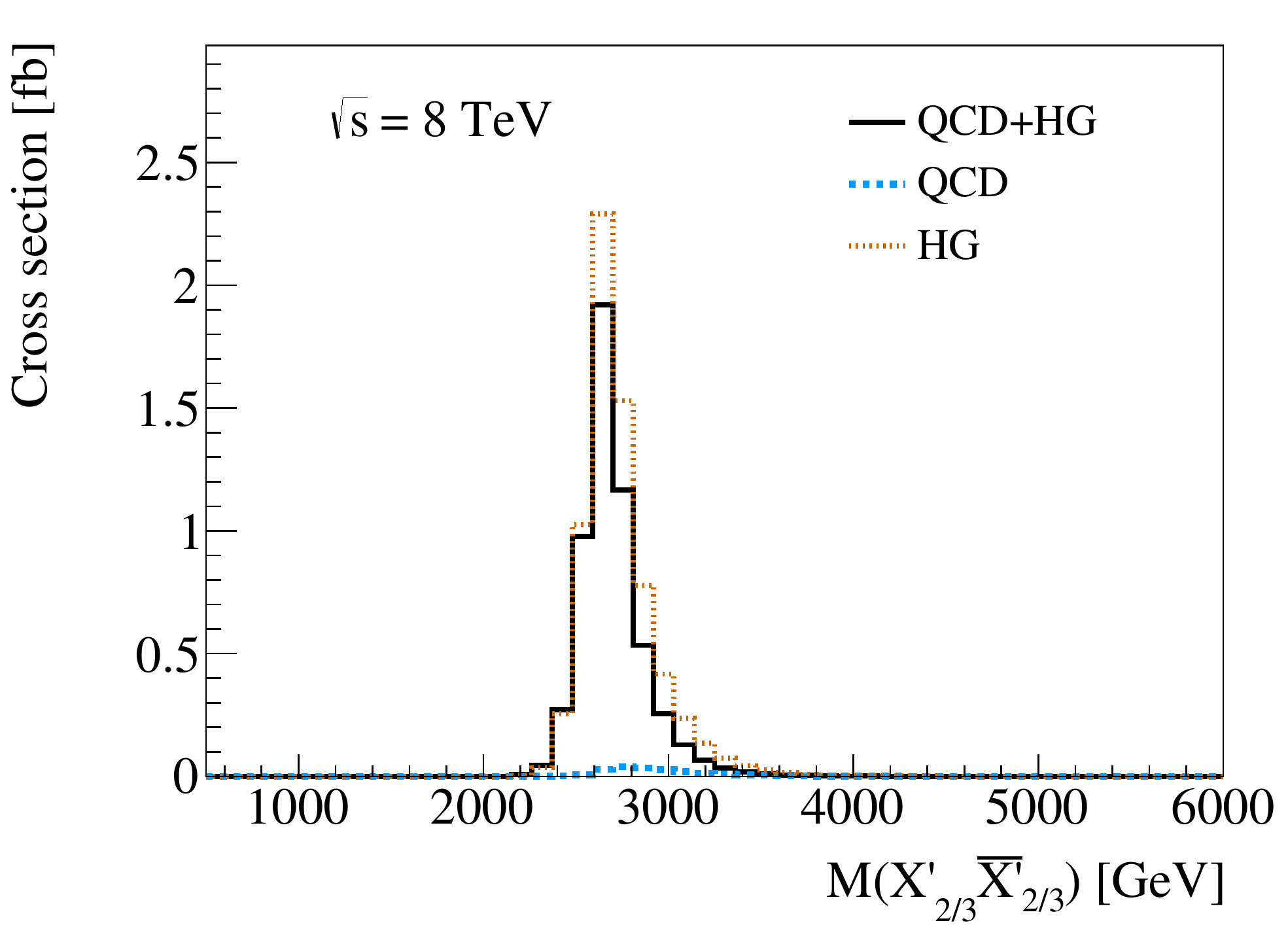}
\caption{}
\end{subfigure}
\begin{subfigure}[b]{0.4\textwidth}
\includegraphics[width=\textwidth]{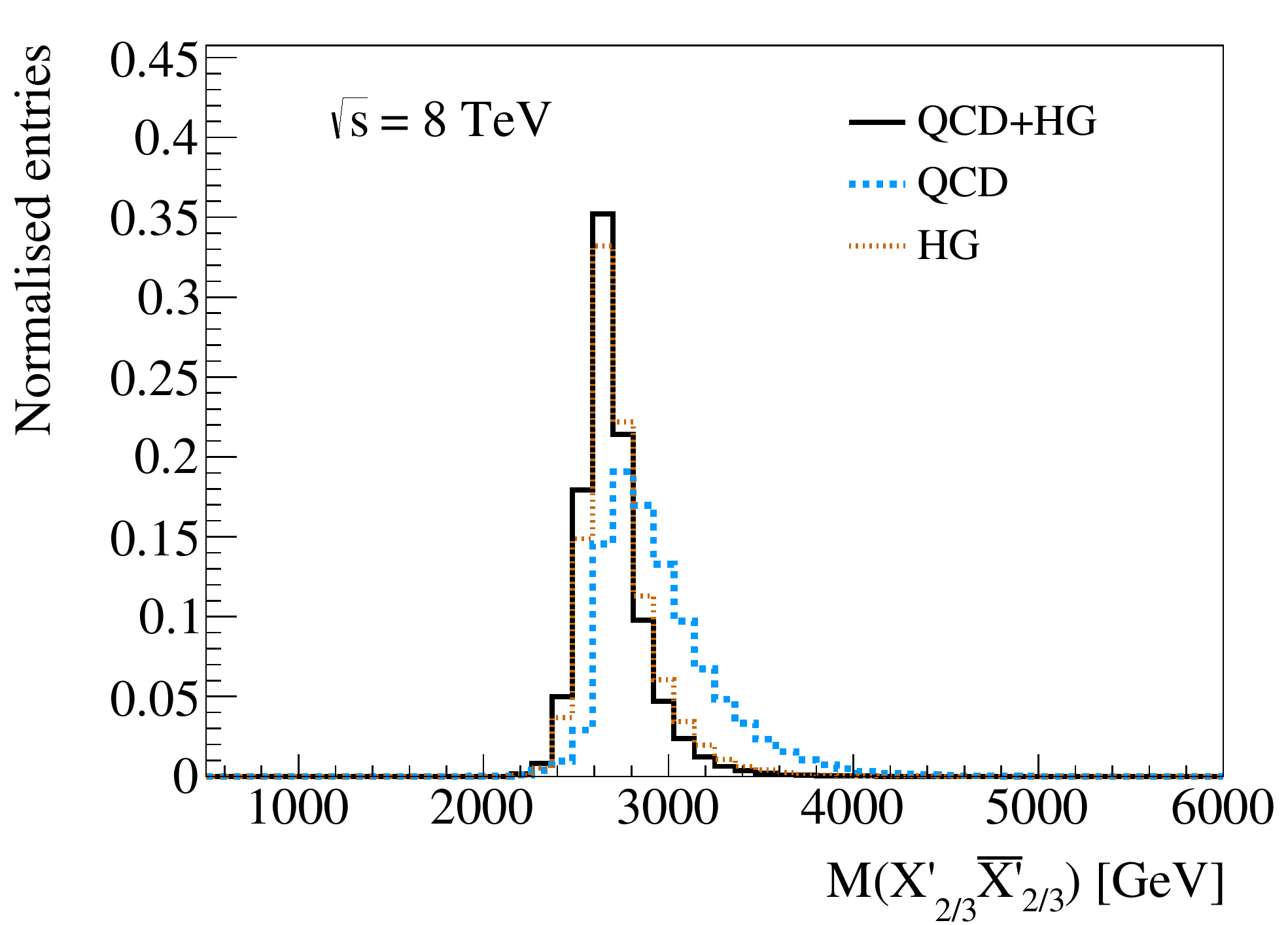}
\caption{}
\end{subfigure}
\end{center}
\caption{$\Xpr$ transverse momentum (top) and $\Xpr \Xprb$
invariant mass (bottom) distributions for $\sqrt{s}=8$~TeV with
$M_G=2.5$ TeV and $M_{\Xpr}=1.3$~\TeV. The left plots are normalized
to the corresponding cross sections whereas the right ones are
normalized to unit area.}
\label{fig:closer8TeV}
\end{figure}

Before considering detector simulation we would like to point out that
the presence of a heavy gluon does not always imply a harder spectrum
than in QCD production. In the plots above, we chose the $M_G=3.5$ TeV
and $M_{\Xpr}=1$ TeV to have a
comparable contribution from QCD and HG productions with a VLQ mass
compatible with current experimental bounds. In this example the heavy
gluon 
mass is large enough to allow the VLQ to always be produced
on-shell. An interesting situation occurs when the heavy gluon mass is
slightly below the kinematical threshold for the VLQ pair production and
the HG production dominates over the QCD one. In this
case, there is still a sizeable production of VLQ pairs in which at least one
of the two VLQ is slightly off-shell, thus making the
corresponding spectrum \textit{softer} than the one in QCD
production. An example of this can be seen in Fig.~\ref{fig:closer8TeV}
in which we show the $p_{\rm T}(\Xpr)$ (top) and 
$M_{\Xpr\Xprb}$ (bottom) distributions for $M_{G}=2.5$~TeV and
$M_{\Xpr}=1.3$ TeV at $\sqrt{s}=8$~TeV. As usual the left plots are
normalized to their corresponding cross sections and the right ones to
unit area. We see in the figure that the heavy gluon resonance slightly below
the pair production threshold makes the spectrum softer. As a result
of this the corresponding analysis efficiencies are smaller for
the QCD+HG production than the ones for QCD. This effect continues at
$\sqrt{s}=13$~TeV despite the larger available phase space.

\subsection{Kinematical differences after detector
  simulation\label{reconstruction}}
  
\begin{figure}[t]
\begin{center}
\begin{subfigure}[b]{0.4\textwidth}
\includegraphics[width=\textwidth]{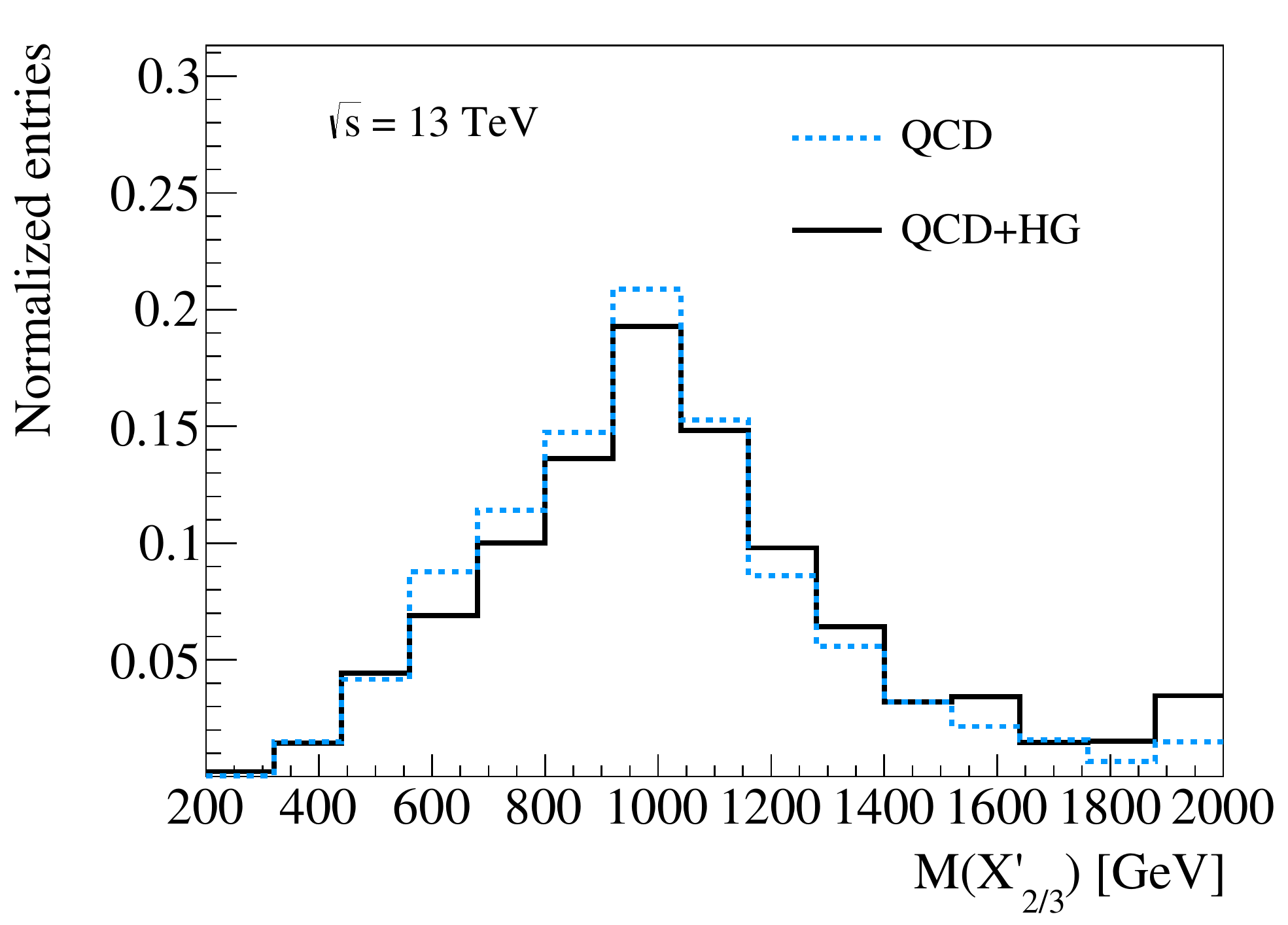}
\caption{}
\end{subfigure}
\begin{subfigure}[b]{0.4\textwidth}
\includegraphics[width=\textwidth]{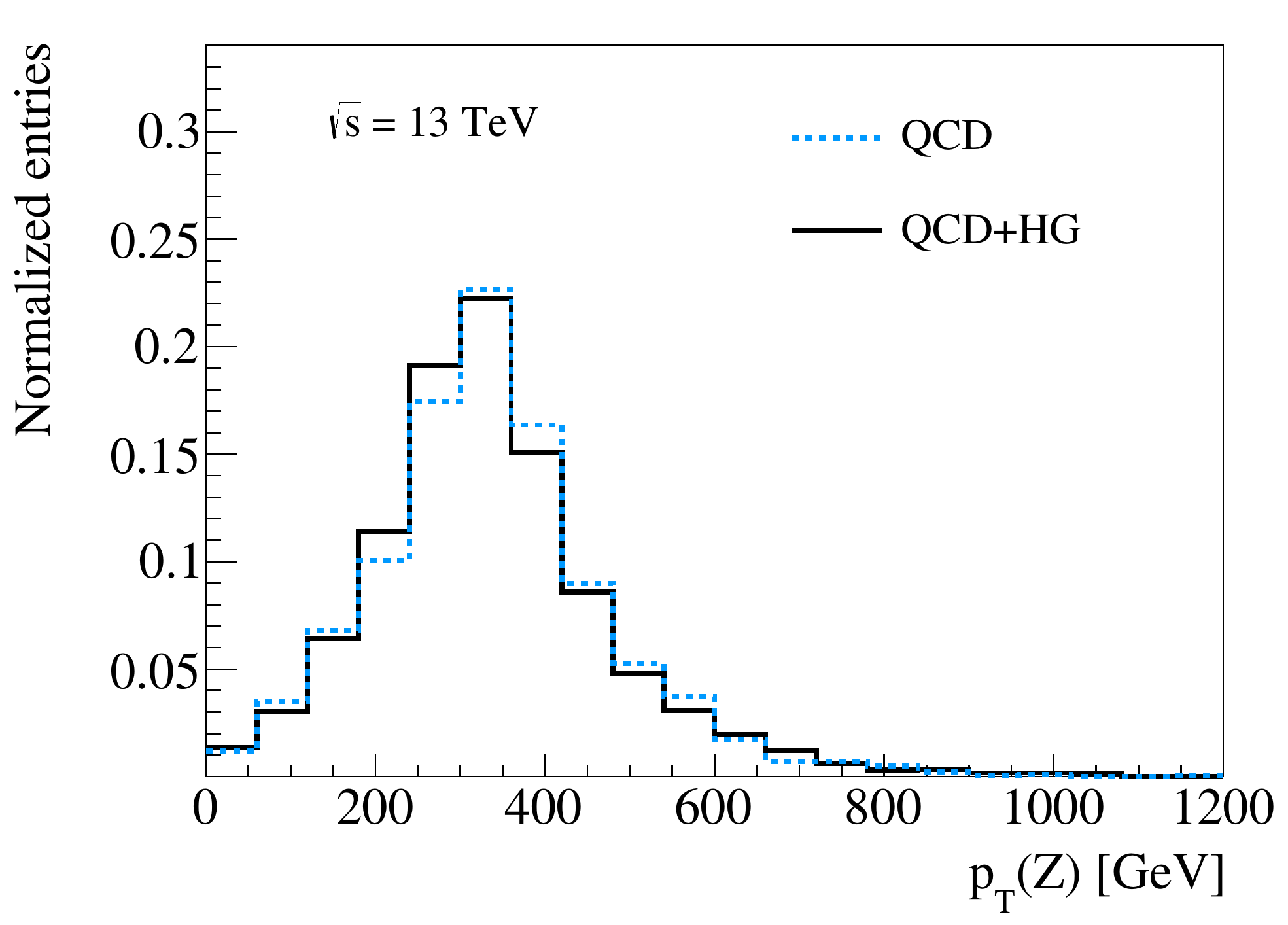}
\caption{}
\end{subfigure}\\
\begin{subfigure}[b]{0.4\textwidth}
\includegraphics[width=\textwidth]{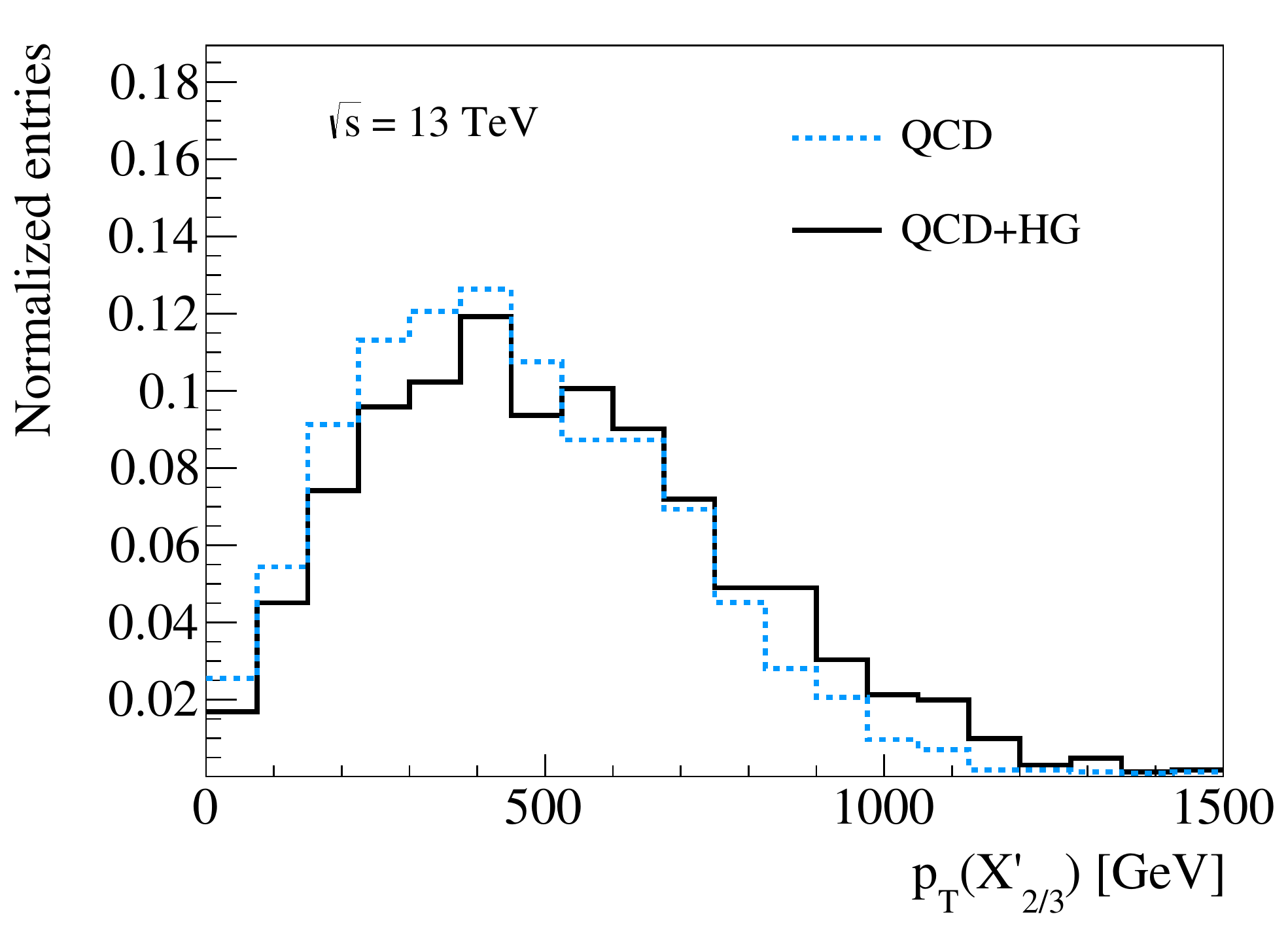}
\caption{}
\end{subfigure}
\begin{subfigure}[b]{0.4\textwidth}
\includegraphics[width=\textwidth]{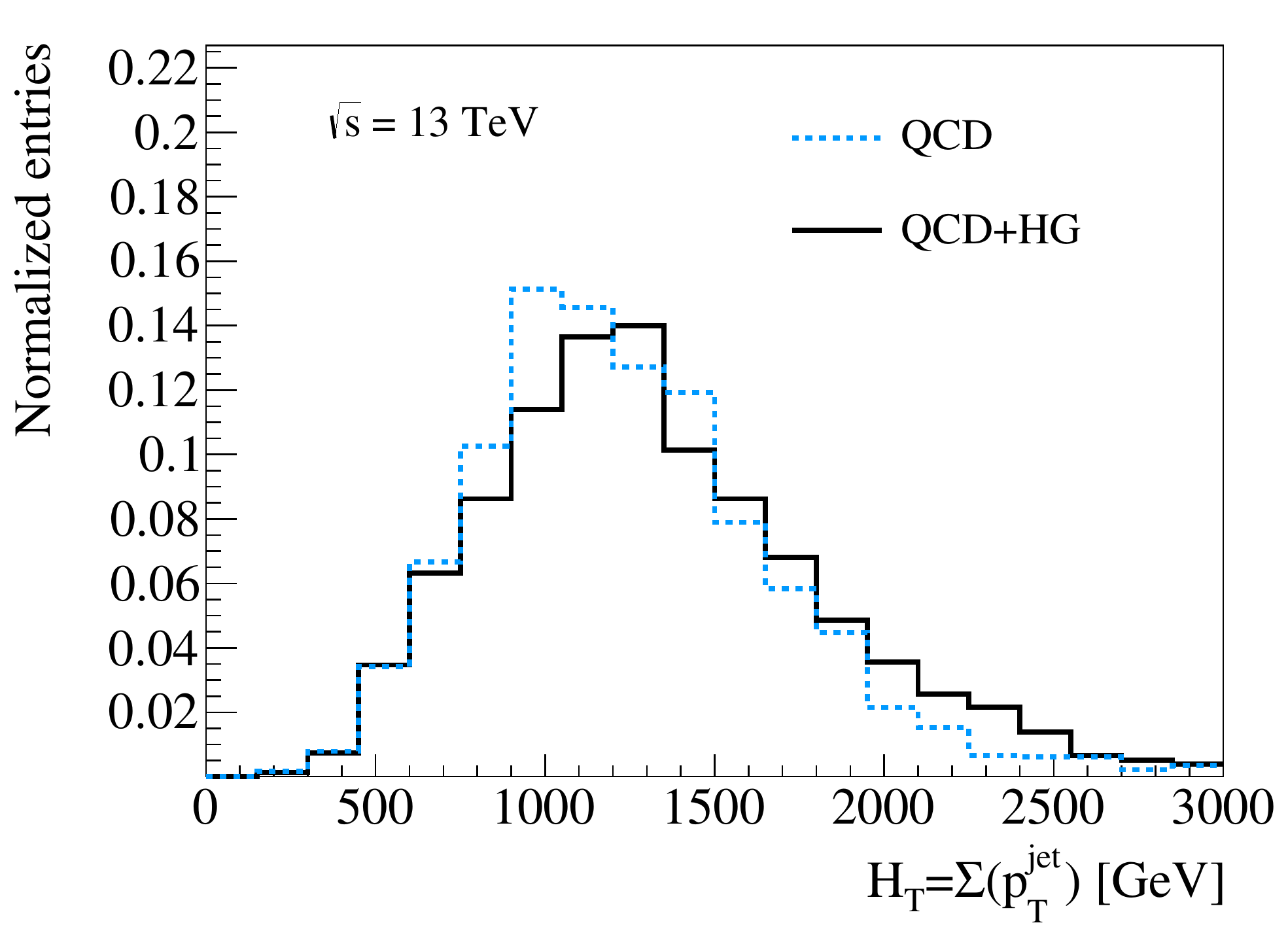}
\caption{}
\end{subfigure}
\\
\end{center}
\caption{
Kinematical distributions after detector simulation for $\sqrt{s}=13$
TeV with $M_{G} = 3.5$~TeV and $M_{\Xpr}=1$~TeV. The distributions
shown are the invariant mass of the objects reconstructing $\Xpr$ (top
left), the $Z$ boson transverse momentum (top right), the
reconstructed $\Xpr$ transverse momentum (bottom left) and
$H_\mathrm{T}$ (bottom right). All distributions are normalized to
unit area.}
\label{fig:delphes}
\end{figure}

Once we have explored the sensitivity of different kinematical
distributions to the presence of a heavy gluon at the parton level, we consider
now a more realistic set-up in which showering/hadronization and
detector simulation are included. 
In order to produce the kinematical distributions after detector
simulation the physics objects need to be properly reconstructed. We
will describe our reconstruction method in detail in the next section
and report here simply the corresponding kinematical distributions. 
Our goal is to show how the
distributions used as example in the previous section get modified
when detector simulation is included. However, aiming at the
recostruction of the full $\Xpr \Xprb$ system is too ambitious if
one is interested in run-1 or early run-2 data. Thus, instead of the
$\Xpr \Xprb$ invariant mass distribution, we will show
another distribution that naturally appears at the detector level,
namely the scalar sum of the transverse momenta of all jets in the
event, $H_{\mathrm{T}}$. We show in Fig.~\ref{fig:delphes} the
invariant mass of the objects that reconstruct the VLQ (top left), to
show that our reconstruction method indeed captures the correct final
state objects; the transverse momentum of the $Z$ bosons (top right);
the transverse momentum of the reconstructed VLQ (bottom left) and
$H_{\mathrm{T}}$ (bottom right). In these plots we have fixed
$M_G=3.5$ TeV, $M_{\Xpr}=1$ TeV and $\sqrt{s}=13$ TeV and we show
only distributions normalized to unit area. The
distributions at $8$ TeV are similar to the ones we show, if anything
with smaller differences between the QCD and QCD+HG productions.
Thus, we see that the somewhat small differences seen at the parton
level are almost completely washed out once detector simulation is
included. This conclusion remains valid for all the kinematical
distributions that we have studied. 
In the case that the heavy gluon is slightly below the
kinematical threshold for pair production and dominates over the QCD
production, in which we saw noticeable shape differences between the
different production mechanism we have checked that these differences
are still visible at the detector level. These kinematical differences
are however overwhelmed by the large enhancement in the cross section
and have little impact on the final limits.

\subsection{Recasting VLQ searches\label{limits:two:ways}}

As we have seen in the previous section, the small kinematical
differences observed at the parton level between the QCD and QCD+HG
production mechanisms are almost completely washed out
once detector simulation is included.
This means that the efficiencies of the experimental analyses, and
therefore the experimental sensitivity, are likely to be very similar
in models with and without a heavy gluon in the spectrum. In this
section we are going to perform a detailed quantitative analysis of
the impact that the small remaining differences have on the final
experimental limits. We will do so by computing the expected limits in
the $M_G-M_{\Xpr}$ plane in two different ways. The first way is to
compute the limits by performing the corresponding analyses on the actual
model with both $\Xpr$ and $G$. The second way is to compute the experimental
sensitivity of the analyses assuming only QCD production, as currently
done by the LHC experimental collaborations. In this second approach
the experimental limit on the corresponding cross section is a
function of $M_{\Xpr}$ only. We will then overlay the theoretical
cross section in our model for different values of $M_G$, assuming
that the efficiencies are identical in both cases and obtaining in
this way the limits as a function of $M_G$. The first approach
corresponds to the properly computed limits, specific for the model at
hand. The second approach, on the other hand, relies on the
(reasonable, as we have seen) assumption that the experimental
efficiencies are quite insensitive to the presence of the HG but has
the advantage that it uses the experimental information that LHC
collaborations are currently publishing and simply rescaling the
theoretical production cross section.

In order to fully exploit the small kinematical differences observed
after detector simulation we will apply the procedure described above
to two different experimental analyses. The first one is a recast of a
search for pair production of
VLQ with at least one of them decaying into a $Z$
published by the ATLAS collaboration~\cite{Aad:2014efa}, 
in which the VLQ is not fully
reconstructed. The second one is a more sophisticated
multivariate analysis, to take full advantage of the small kinematical
differences, in which the VLQ if fully reconstructed.
Let us discuss the two analyses and the corresponding limits in turn.
\begin{table}[t]
\begin{center}
\begin{tabular}{|l|c|c|}
\hline
\textbf{Selection} & \textbf{QCD cut efficiency (\%)} & \textbf{QCD+HG cut efficiency (\%)} \\ \hline
Leptonic $Z$ & 1.28 & 1.36\\ \hline
$\geq 2$ jets & 99.82 & 99.90 \\ \hline
$\geq 2$ $b$-jets & 64.25 & 64.00 \\ \hline 
$p_{\rm T}(Z) > 150$~GeV & 93.09 & 92.50 \\ \hline 
$H_{\rm T} > 600$~GeV & 94.19 & 93.42 \\ \hline 
\end{tabular}
\end{center}
\caption{Event selection and cut efficiencies for both QCD and QCD+HG
productions at $\sqrt{s}=8$~TeV for $M_G=3.5$~TeV and $M_{\Xpr}=1$~TeV.
Each efficiency is derived based on the number
of events which passed the cut before. The efficiency of the first cut
includes the leptonic BR of the $Z$.}
\label{tab:selection}
\end{table}

In our first analysis we have replicated the ATLAS selection for the
dileptonic channel 
which is summarized in Table~\ref{tab:selection}, together with the
efficiency of each cut 
at $8$~TeV for $M_G=3.5$ TeV and $M_{\Xpr}=1$ TeV.
First, a pair of opposite-sign same-flavour leptons 
($p_T(\ell) > 25$~GeV, $|\eta(e)|< 2.47$, excluding electrons within $1.37 < |\eta| < 1.52$, and $|\eta(\mu)| < 2.5$) are
required with $|m_{\ell\ell}-m_{Z}|<10$~GeV. The events passing the
first cut are required to have at least $2$ jets ($p_T(j)> 25$ GeV, $|\eta(j)|<5$), with at least two of
them being $b$-tagged.
Two further cuts are applied to reduce background contamination: 
$p_{\rm T}(Z) > 150$~GeV and $H_{\rm T}> 600$~GeV, where $H_{\rm T}$ 
is the scalar sum of the $p_{\rm T}$ of all selected
jets. As we can see in the table the efficiencies are very similar for
the QCD and QCD+HG productions.
\begin{figure}[t]
\begin{center}
\begin{subfigure}[b]{0.4\textwidth}
\includegraphics[width=\textwidth]{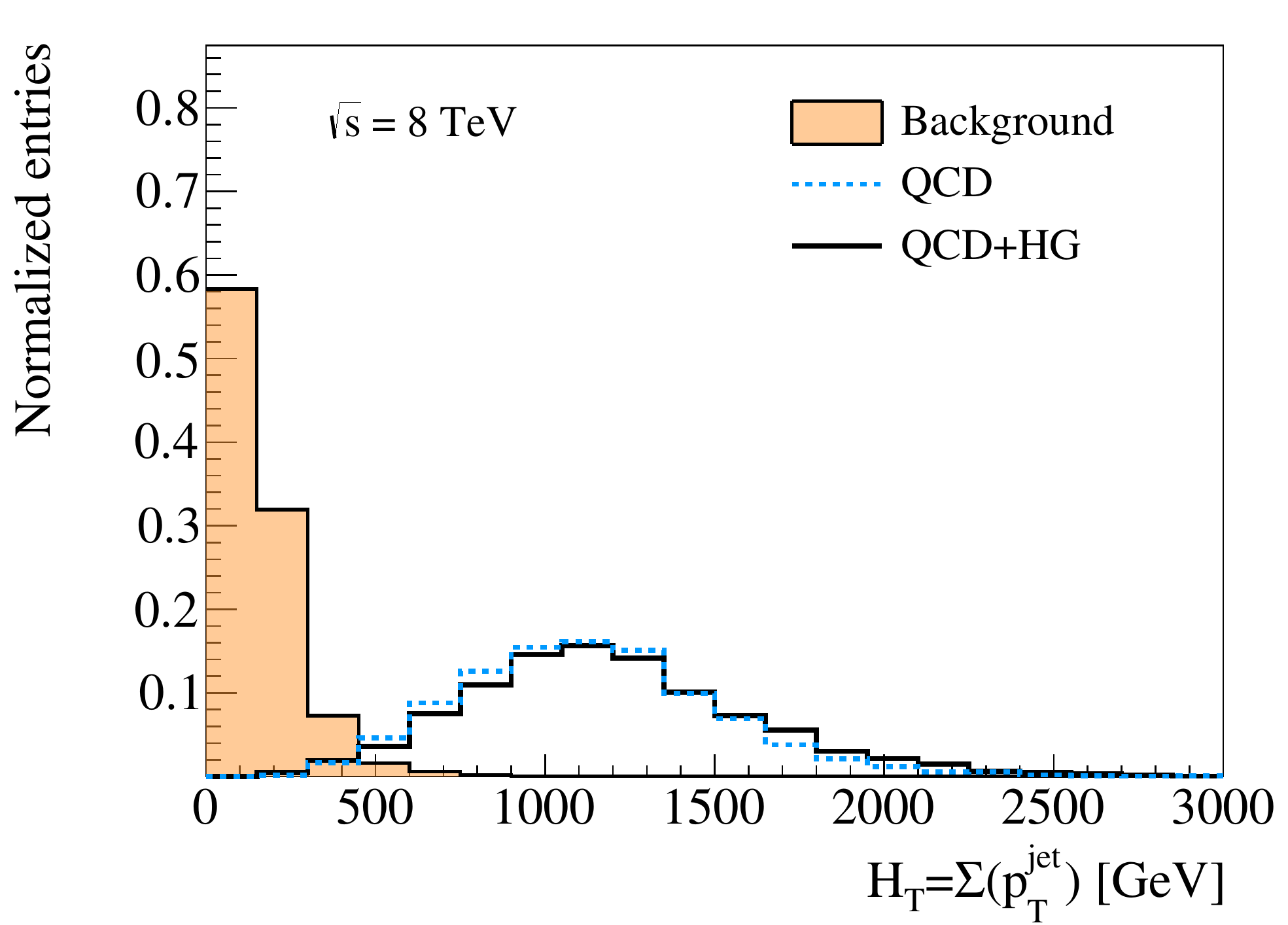}
\caption{}
\end{subfigure}
\begin{subfigure}[b]{0.4\textwidth}
\includegraphics[width=\textwidth]{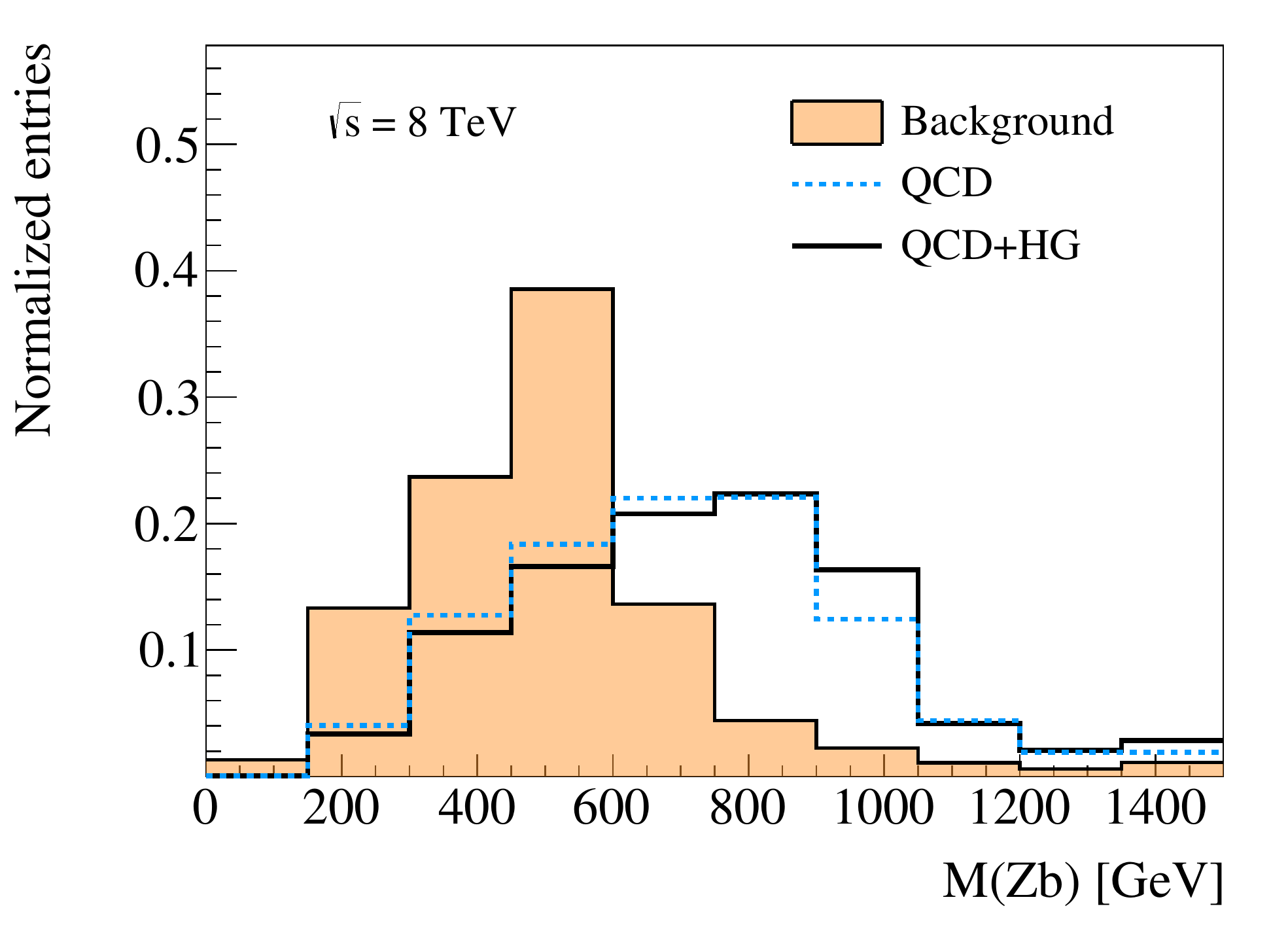}
\caption{}
\end{subfigure}\\
\end{center}
\caption{Signal and background distributions at $\sqrt{s}=8$~TeV corresponding
to $H_{\rm T}$ (left) for events passing the first three cuts and and
the invariant mass of the $Zb$ system applying all selection (right).
The distributions are normalized to unit area. We have fixed $M_G=3.5$
TeV and $M_{\Xpr}=1$ TeV.}
\label{fig:ATLASdist8TeV}
\end{figure}
We have also simulated the two main backgrounds identified
by the ATLAS analysis, namely $Z$+jets and $t\bar{t}$, using the same
tools as for the signal. 
For the $Z$+jets background we have generated a matched sample of 
$Z$+bb and $Z$+cc with up to 3 additional jets in the matrix element. No 
$Z$+light sample has been generated since after the 2 $b$-jet
selection the contribution is negligible. The $t\bar t$ sample has
been generated at NLO in \texttt{MG5} and the dileptonic decay has been
performed with \texttt{Madspin}~\cite{Artoisenet:2012st}. 

\begin{figure}[t]
\begin{center}
\begin{subfigure}[b]{0.48\textwidth}
\includegraphics[width=\textwidth]{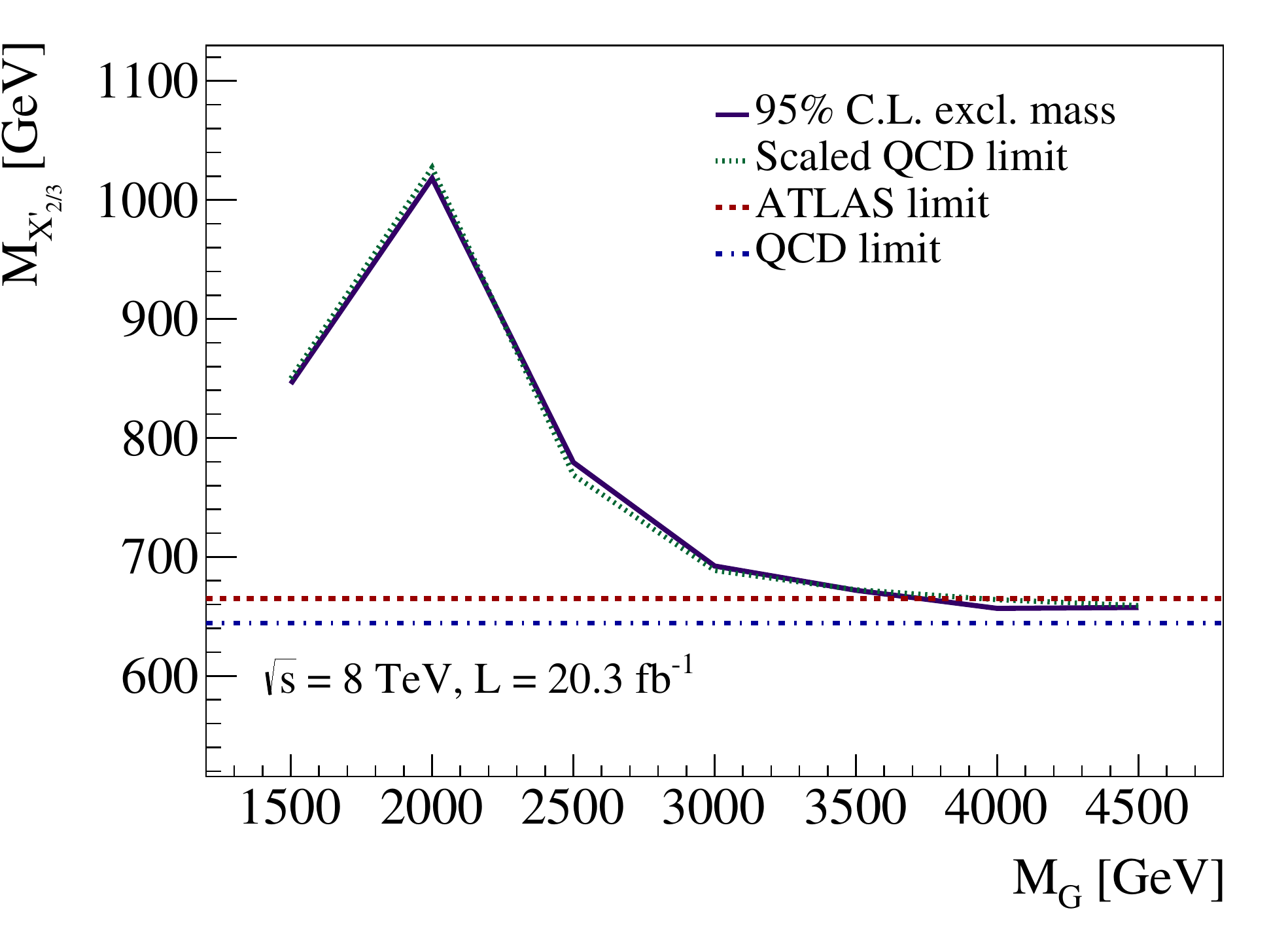}
\caption{}
\end{subfigure}
\begin{subfigure}[b]{0.48\textwidth}
\includegraphics[width=\textwidth]{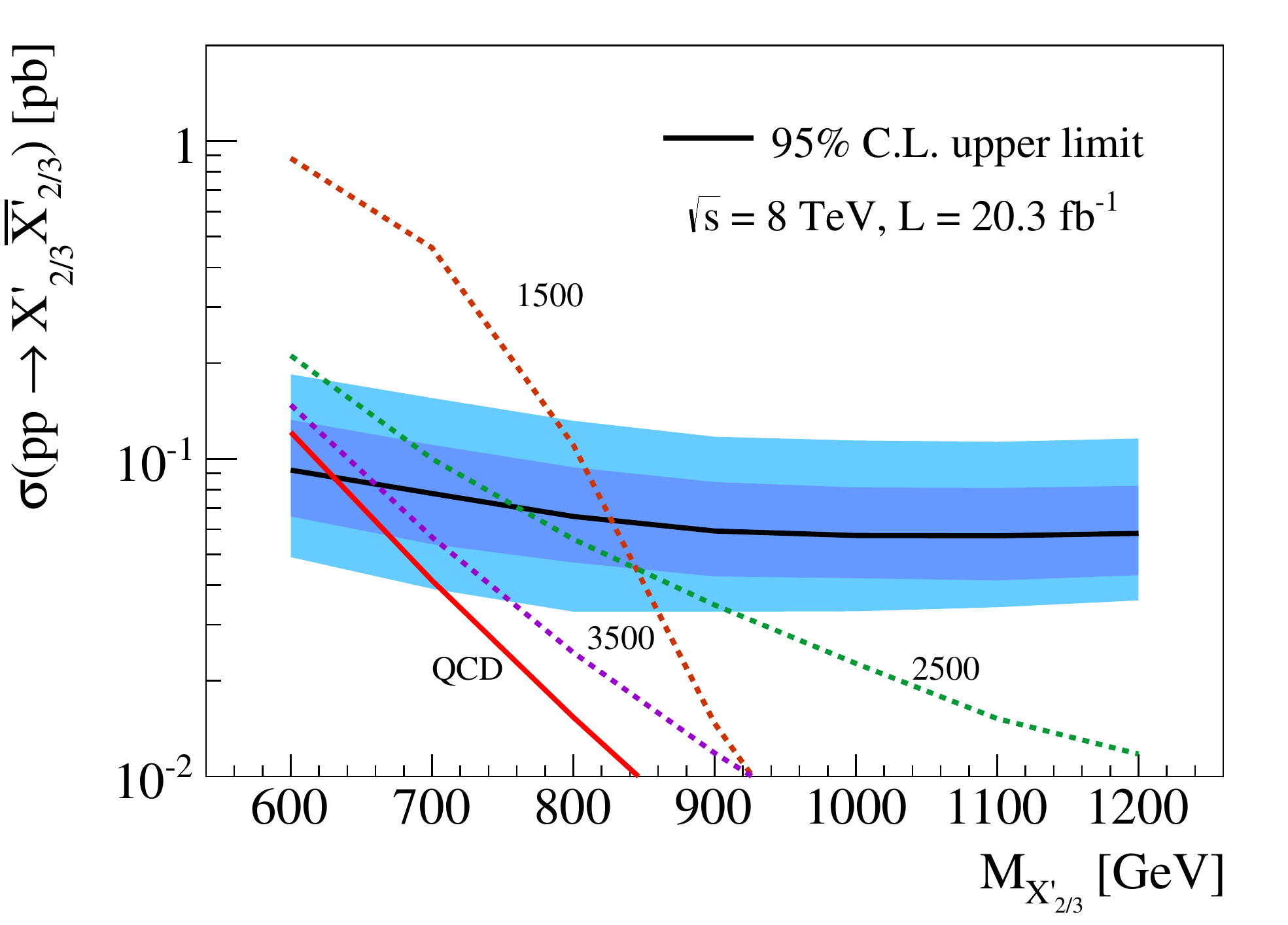}
\caption{}
\end{subfigure}\\
\end{center}
\caption{Left: 95\% confidence level (C.L.) lower limits in the $M_G-M_{\Xpr}$ plane computed with
the two methods discussed in the text. We show for comparison the
limit reported by ATLAS ($\sqrt{s}=8$~TeV) in dashed red and the corresponding number
that we have obtained with our recast of that analysis in dashed
blue. Right: 95\% C.L. upper limits on the
pair production cross section as a function of $M_{\Xpr}$ assuming QCD
production (solid black) with 1- and 2-$\sigma$ bands in blue, together
with the theoretical production cross sections in the model for
different values of $M_G$. }
\label{fig:8TeVZtaglimits}
\end{figure}

As expected the relevant kinematical distributions after cuts present
minimal differences between the QCD and QCD+HG production
mechanisms. As an example we show in 
Fig.~\ref{fig:ATLASdist8TeV} the $H_{\rm T}$ distribution
after the three first cuts (left) and the invariant mass of the $Zb$
system (right), which is the discriminant variable defined as in the ATLAS
publication: the $Z$ boson is the candidate selected before and the
$b$-jet is the highest-$p_{\rm T}$ $b$-tagged jet in each event. We also show
the corresponding background distibutions for comparison. All
distributions are normalized to unit area and we have fixed $M_G=3.5$
TeV and $M_{\Xpr}=1$ TeV.
Finally, we show the corresponding limits in the $M_G-M_{\Xpr}$ plane
computed with the two methods described above in
the left panel of Fig.~\ref{fig:8TeVZtaglimits}. The solid purple line
corresponds to the bounds computed within the model with the HG
included. The green line, labeled ``Scaled QCD limit'', corresponds to the
bounds computed assuming 
that the experimental sensitivity is the same in the QCD and QCD+HG
production mechanisms. This second method is illustrated in the right
panel of Fig.~\ref{fig:8TeVZtaglimits}, in which we show in solid black
the experimental limit on the $\Xpr \Xprb$ production cross
section as a function of $M_{\Xpr}$ assuming QCD production, as well as
the theoretical pair production cross section, assuming QCD+HQ production, for
different values of $M_G$. 
The points in which these curves cross the experimental
limit determine the solid green curve on the left panel.
As we can see in the figure, the two ways of computing the limits give
virtually identical results. This indicates that, at least with
current analyses at 8 TeV that do not fully reconstruct the VLQ, one
can simply use the experimental limits on the production cross
section as a function of the VLQ mass published by the experimental
collaborations to put limits on models with a heavy gluon.
We also show in the left panel with horizontal lines the limit we
obtain in the case of QCD production and the limit reported for that
case by ATLAS. The level of agreement between the two and of them with
the large $M_G$ limit shows the consistency of our results.

Let us now consider the second, more sophisticated, type of analysis to
try to exploit the small kinematical differences observed at the
detector level.
In this analysis we modify slightly the event selection and
reconstruct the $\Xpr$ decay to a $Z$ boson as follows.
The $Z$ boson is reconstructed as in the previous analysis.
We then require at
least $4$ jets with at least 2 of them being $b$-tagged
jets. Two of the selected jets are required to have
an invariant mass inside a $10$~GeV window of
$W$ mass and the two leptons reconstructing the $Z$ boson are required
to satisfy $\Delta R(\ell_1,\ell_2) < 1.5$.
These two last cuts reduce the background contamination while still
leaving the $p_{\rm T}$ and $H_{\rm T}$ variables with discriminant
power. For the reconstruction of the $W$, top and $\Xpr$ we use a
$\chi^2$ to find the best combination of objects reconstructing the
desired particles.
Specifically, for a particle $X$ which decays into two
particles $i,j$, we define $\bar\chi^2_X$ as 
\begin{equation}
\bar\chi^2_X = \frac{(M_{ij}-M_{X})^2}{\sigma_X^2}\Delta R(i,j),
\end{equation}
where $M_{ij}$ is the invariant mass of the $i,j$ system, $M_X$ is the
mass of the mother particle, $\sigma_X$ is the width of the mother
particle and $\Delta R(i,j)$ is the angular separation between both
decay products. Using this definition we select the jets and $b$-jet
pairing that minimizes  
\begin{equation}
\left( \bar\chi^2_W + \bar \chi^2_t\right)\Delta R(Z,t).
\end{equation}
Only hadronic decays of the $W$ boson are considered and the top quark
is reconstructed with the reconstructed $W$ boson and a $b$-jet. The
$\Delta R$ term in the $\bar\chi^2_X$ definition is used to include
the angular separation between the decay products in the object
reconstruction. 

\begin{figure}[ht!]
\begin{center}
\begin{subfigure}[b]{0.48\textwidth}
\includegraphics[width=\textwidth]{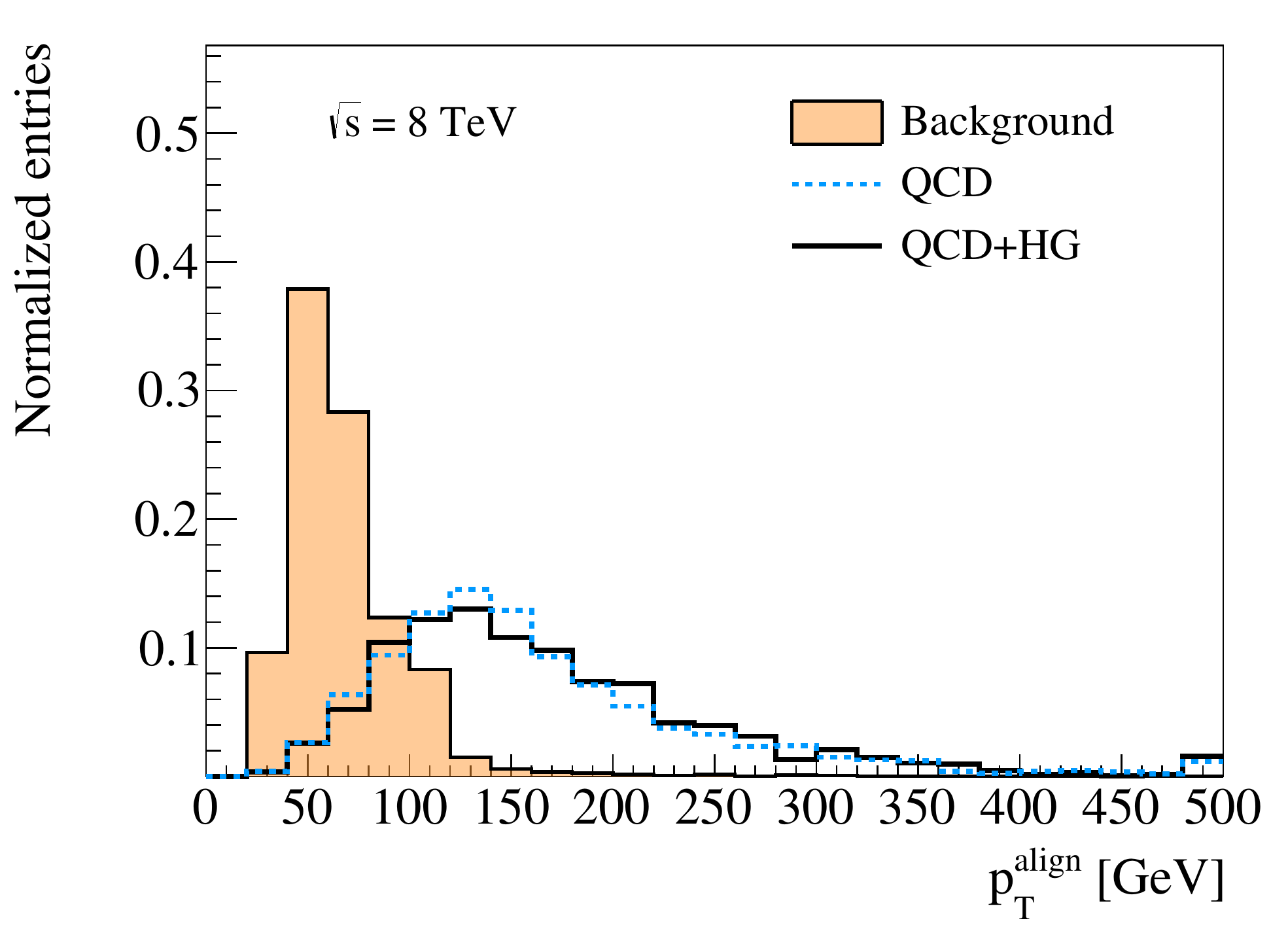}
\caption{}
\end{subfigure}
\begin{subfigure}[b]{0.48\textwidth}
\includegraphics[width=\textwidth]{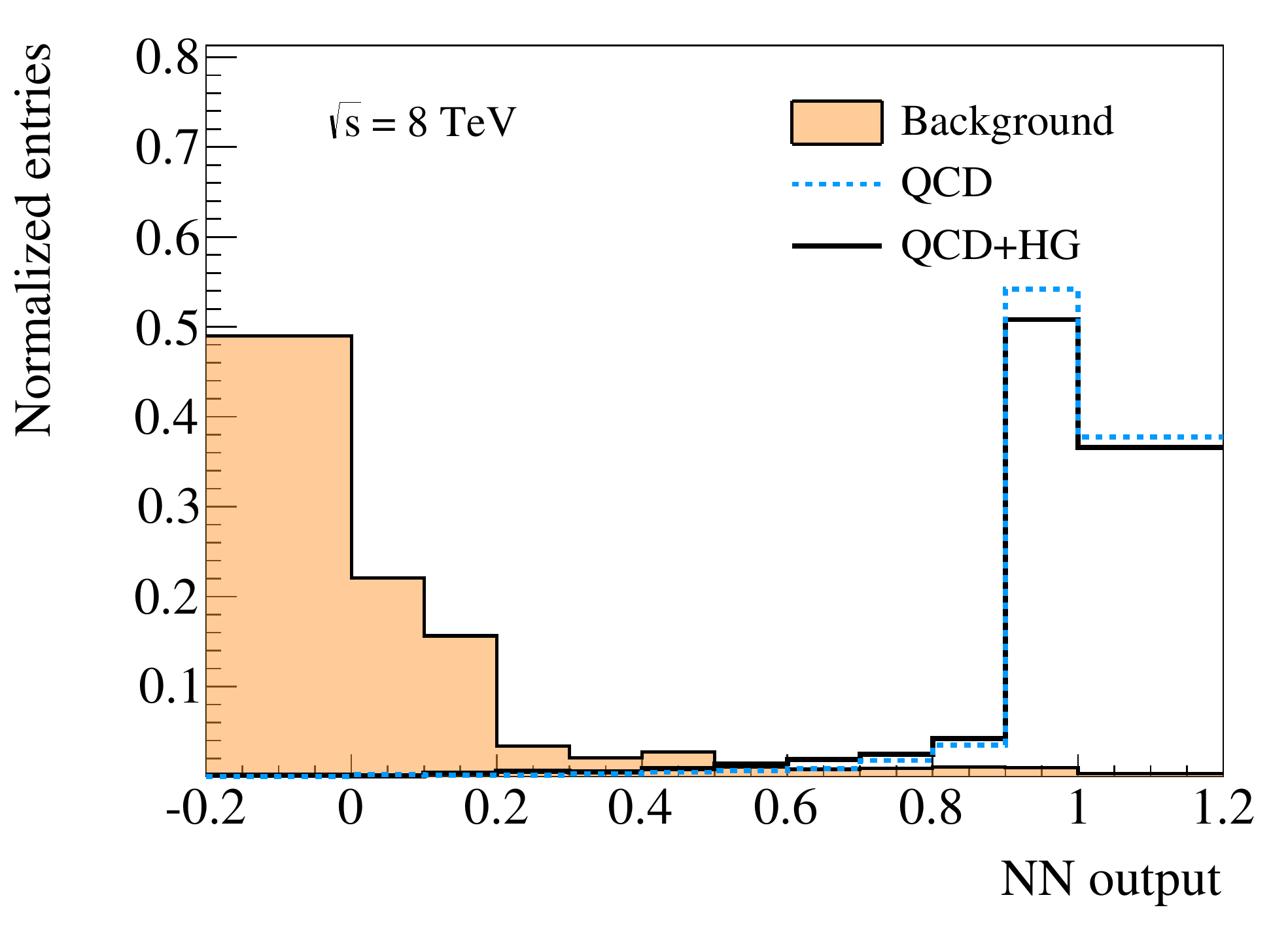}
\caption{}
\end{subfigure}\\
\end{center}
\caption{$p_{\rm T}^{align}$ (left) and NN output (right)
distributions for signal and background normalized to unit area
for $\sqrt{s}=8$ TeV, $M_{G} = 3.5$~TeV 
and $M_{\Xpr} = 1$~TeV.} 
\label{fig:tmva8TeV}
\end{figure}

After this selection is applied we define a multivariate analysis
using a neural network (NN) implemented through {\tt
TMVA}~\cite{Hocker:2007ht} with 7 input distributions: Jet
multiplicity, $p_{\rm T}(t)$, $p_{\rm T}(\Xpr)$, $p_{\rm T}(Z)$,
$H_{\rm T}$, $M_{\Xpr}$ and $p_{\rm T}^{align}$. 
$p_{\rm T}^{align}$ is a variable that 
is sensitive to both the energy of the decay
products of the vector-like quark and the angular separation between
them. It increases as we move to a more boosted regime and
is defined as 
\begin{equation}
p_{\rm T}^{align} = \frac{\sqrt{p_{\rm T}(Z)^2+p_{\rm T}(W)^2+p_{\rm
T}(b)^2}}{\mathrm{max}
\left(\Delta R(\Xpr,Z),\Delta R(\Xpr,W), \Delta R(\Xpr,b)\right)}.
\end{equation}
These variables are selected to provide good discrimination between
signal and background and for their potential to provide discrimination
between both production mechanisms (QCD and QCD+HG). The NN is trained
for each pair of masses of $G$ and $\Xpr$ quark separately and
the output of the NN is used as discriminant variable to derive
exclusion limits. We show in Fig.~\ref{fig:tmva8TeV} the signal and
background distributions of $p_{\rm T}^{align}$ (left) and the NN
output (right) for $\sqrt{s}=8$ TeV, $M_{G} = 3.5$~TeV 
and $M_{\Xpr} = 1$~TeV. 
In both cases the distributions are normalized to unit area.

\begin{figure}[ht!]
\begin{center}
\begin{subfigure}[b]{0.6\textwidth}
\includegraphics[width=\textwidth]{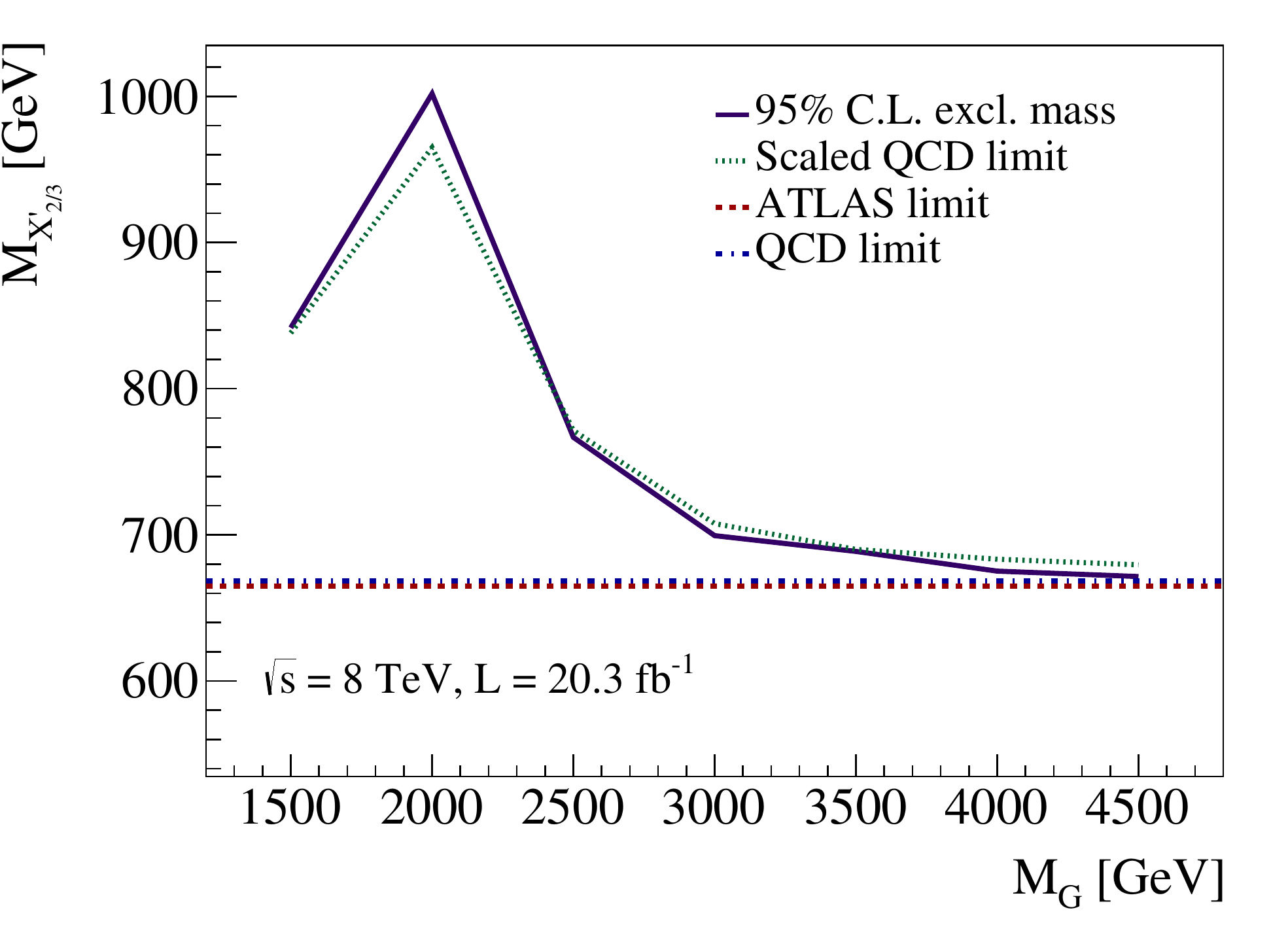}
\caption{}
\end{subfigure}
\end{center}
\caption{95\% C.L. lower limit on $M_{\Xpr}$ as a function of $M_G$
for $\sqrt{s}=8$~TeV with the full run-1 integrated luminosity. 
The green line shows the lower
mass limit assuming that the only effect is an increase in the cross
section and the purple line shows the lower mass limit when
considering the correct cross section upper limit for each pair of
masses.} 
\label{fig:8TeVtmvalimits}
\end{figure}

We use the $CL_s$ method~\cite{Read:2002hq} with the NN output as the discriminating
variable to compute the limits in the $M_G-M_{\Xpr}$ plane with the
two methods described at the beginning of this section. 
The result is shown in Fig.~\ref{fig:8TeVtmvalimits} using the same
color coding as in the left panel of Fig.~\ref{fig:8TeVZtaglimits}.
We see from the results in the figure that, although the multivariate
analysis seems to be more sensitive to the presence of the heavy
gluon, the difference between the properly computed limit, taking into
account the presence of HG, and the one in which the QCD experimental
limit is used is only marginal. The improvement in the actual limits
is also minimal. Thus, the conclusion drawn from the previous analysis
that one can simply take the experimentally published limits that
assume QCD production and use them to compute the bounds on models
with a heavy gluon seems to be robust and apply also to more
sophisticated analyses with a richer object reconstruction.

\section{Limits on heavy gluons from vector-like quark searches\label{limits}}

Once we have checked in detail that the efficiencies and therefore the
experimental limits on the cross section of pair produced VLQ 
are insensitive to the
presence of the heavy gluon we proceed to compute the limits in the
$M_G-M_{Q}$ plane for our CHM, where now $Q$ stands
for any of the VLQ present in the spectrum. In
Section~\ref{sec:current_limits} we use the published run-1 limits on
VLQ pair production (QCD only) to compute the current bounds on the model
assuming the experimental efficiencies to be insensitive to the
presence of the heavy gluon. We then proceed in
Section~\ref{sec:early_run_2} to estimate the bounds that can be
obtained from the multivariate analysis described in the previous
section with the early run-2 data (we assume an integrated luminosity
of 10 fb$^{-1}$).

\subsection{Current limits\label{sec:current_limits}}

In this section we reinterpret the current results published by the
ATLAS and CMS collaborations assuming that the experimental
efficiencies are insensitive to the presence of the heavy gluon. 
We have tried to be as complete as possible using different analyses
that target all the VLQ present in the model
under study. 

When trying to use the published limits we find a first obstacle. We have
checked in the previous sections the independence of the efficiencies
to the presence of the heavy gluon, 
however they can be quite sensitive to the decay
BR assumed for the VLQ. The reason is
cross-contamination between different decay channels. A specific
analysis is more sensitive to certain decay channels than others and
therefore if the BR change we change the population
of samples with different efficiencies and thus the global efficiency
of the analysis. We have used a specific
analysis whenever the BR assumed for the channel that that particular
analysis is most sensitive to is equal or smaller than the one in our model.
This gives us a conservative bound since we would be assuming in this
way a global efficiency that is equal or smaller than the one in our
model, due to the contamination of channels with smaller efficiencies.
The experimental collaborations have made a tremendous effort in
obtaining lower limits on the mass of the VLQ for arbitrary BR. In
doing that, they have computed the corresponding limits in the
production cross section as a function of the VLQ mass. It would be
extremelly useful to provide that information too, which would allow
us to compute the limits in models with heavy gluons using the correct BR.

\begin{figure}[t]
\begin{center}
\begin{subfigure}[b]{0.48\textwidth}
\includegraphics[width=\textwidth]{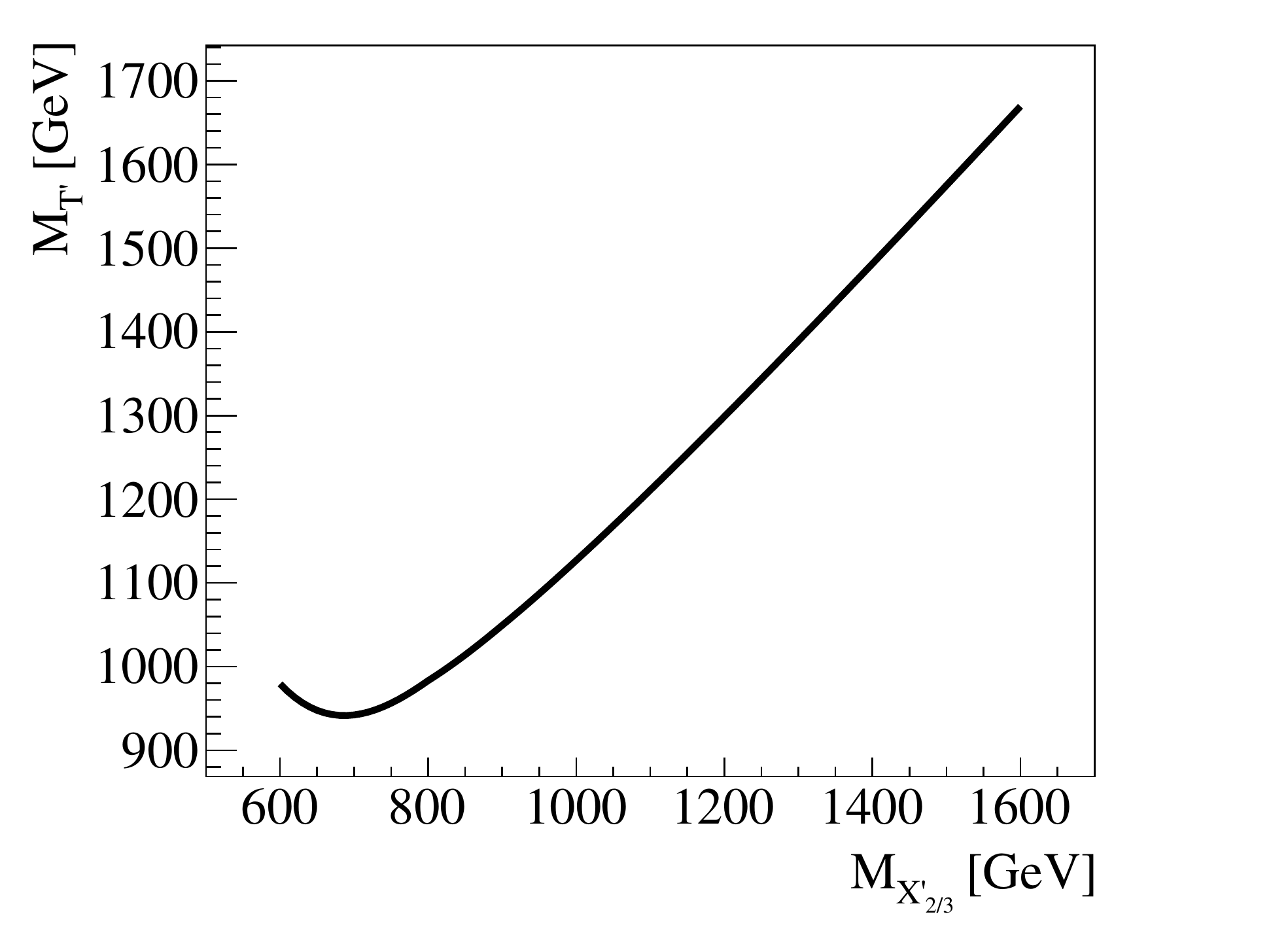}
\caption{}
\end{subfigure}
\begin{subfigure}[b]{0.48\textwidth}
\includegraphics[width=\textwidth]{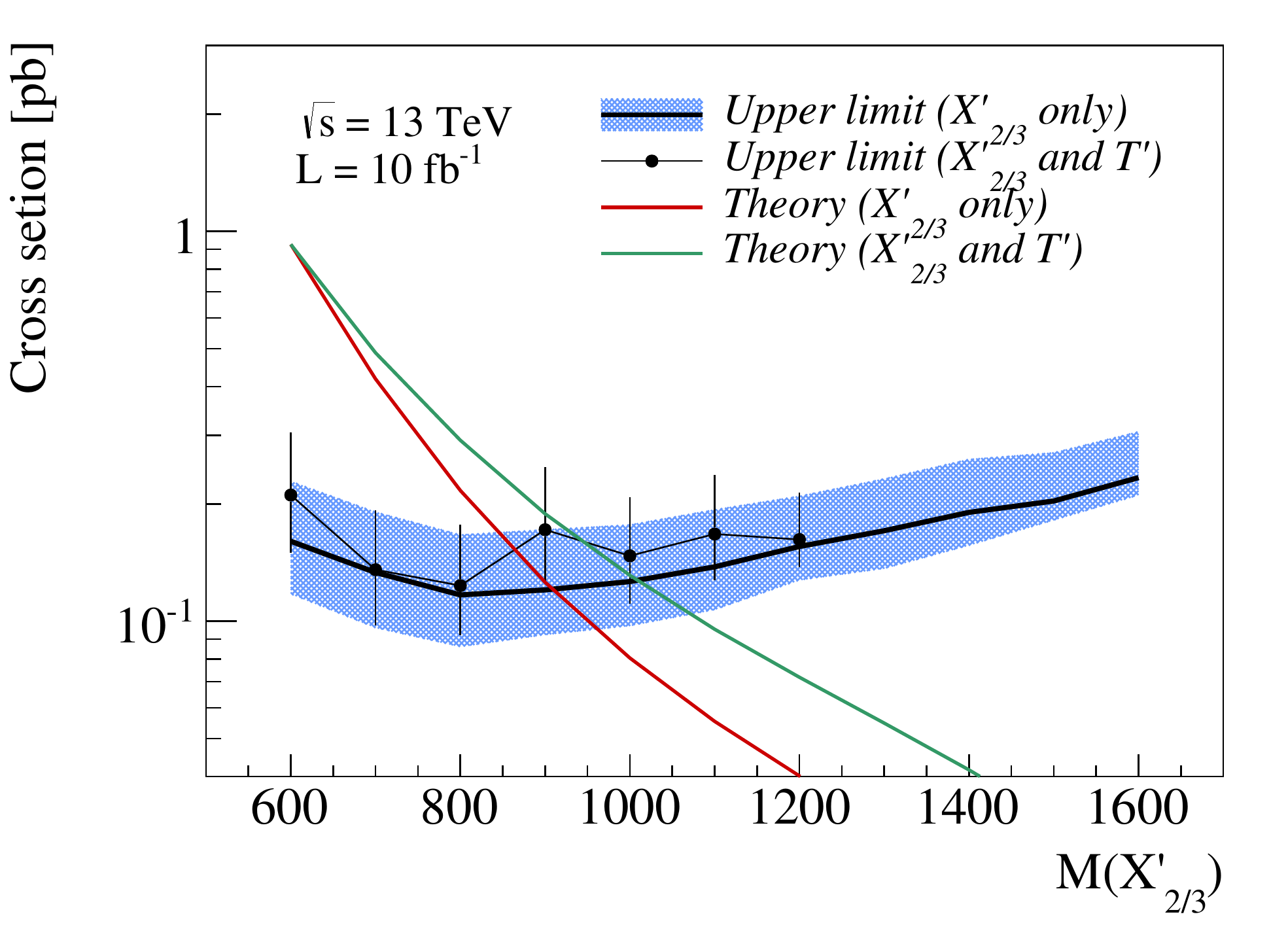}
\caption{}
\end{subfigure}
\end{center}
\caption{Left: $M_{T^\prime}$ mass as a function of $M_{\Xpr}$ mass for the
chosen parameters. 
Right: comparison between the 95\%{} C.L. experimental limit 
obtained assuming 
only $\Xpr$ production (black solid line) and assuming production of
both $\Xpr$ and $T^{\prime}$ (black dots). The solid red line
represents the cross section of pair production of $\Xpr$ only while
the dashed green line represent the cross section of pair production
of both $\Xpr$ and $T^{\prime}$. In this figure $M_{G} = 3.5$~TeV 
and $\sqrt{s} = 13$~TeV have been chosen as an
illustrative example. Similar behaviour is found for different masses
and center-of-mass energy.} 
\label{fig:ht12compare}
\end{figure}

A second difficulty arises from the fact that the experimental limits
are always reported assuming that only one VLQ is present 
in the spectrum. In our
case, due to the specific BR pattern, we can group the VLQ in two
different groups, $X_{5/3}$ and $B$ both decay into $Wt$ and,
unless full reconstruction of one decay leg with same/opposite sign
leptons is performed, they are difficult to distinguish
(see~\cite{AguilarSaavedra:2009es}
for studies on how to characterize the different top
partners in a similar model, assuming QCD production). 
A second group is formed by $\Xpr$ and $T^\prime$, which have the
exact same decay patterns. In our model, $X_{5/3}$ and $\Xpr$ are
degenerate and lighter than $T^\prime$ and $B$. 
In the region of parameter space that we are exploring the
mass splitting is of the order of 10\%, which is large enough to make
interference effects small but not so large that two distinct peaks
would show up in the relevant distributions. Furthermore, the heavier
ones, $T^\prime$ and $B$, are heavy enough to make their 
contribution to the total
cross section much smaller than the one from the lighter ones,
$X_{5/3}$ and $\Xpr$, in the searches to which both contribute. We
show in the left panel of Fig.~\ref{fig:ht12compare} the $T^\prime$
mass as a function of the $\Xpr$ mass, for the chosen parameters.
Thus,
the global efficiencies are likely to be quite insensitive to their
presence.
In order to be quantitative, we have repeated
our multivariate analysis including the two charge $2/3$ quarks, 
$\Xpr$ and $T^\prime$, for a
few selected points. We show in the right panel of 
Fig.~\ref{fig:ht12compare} 
a comparison between the experimental sensitivity
(limit on the production cross section) obtained in the full model
with both quarks and the one obtained neglecting the presence of
$T^\prime$. As we can see the two curves are compatible within
uncertainties and 
we can therefore compute the limits overlaying the total cross section
(including $\Xpr$ and $T^\prime$) over the published experimental
bounds on the production cross section.

\begin{figure}[ht!]
\begin{center}
\begin{subfigure}[b]{0.49\textwidth}
\includegraphics[width=\textwidth]{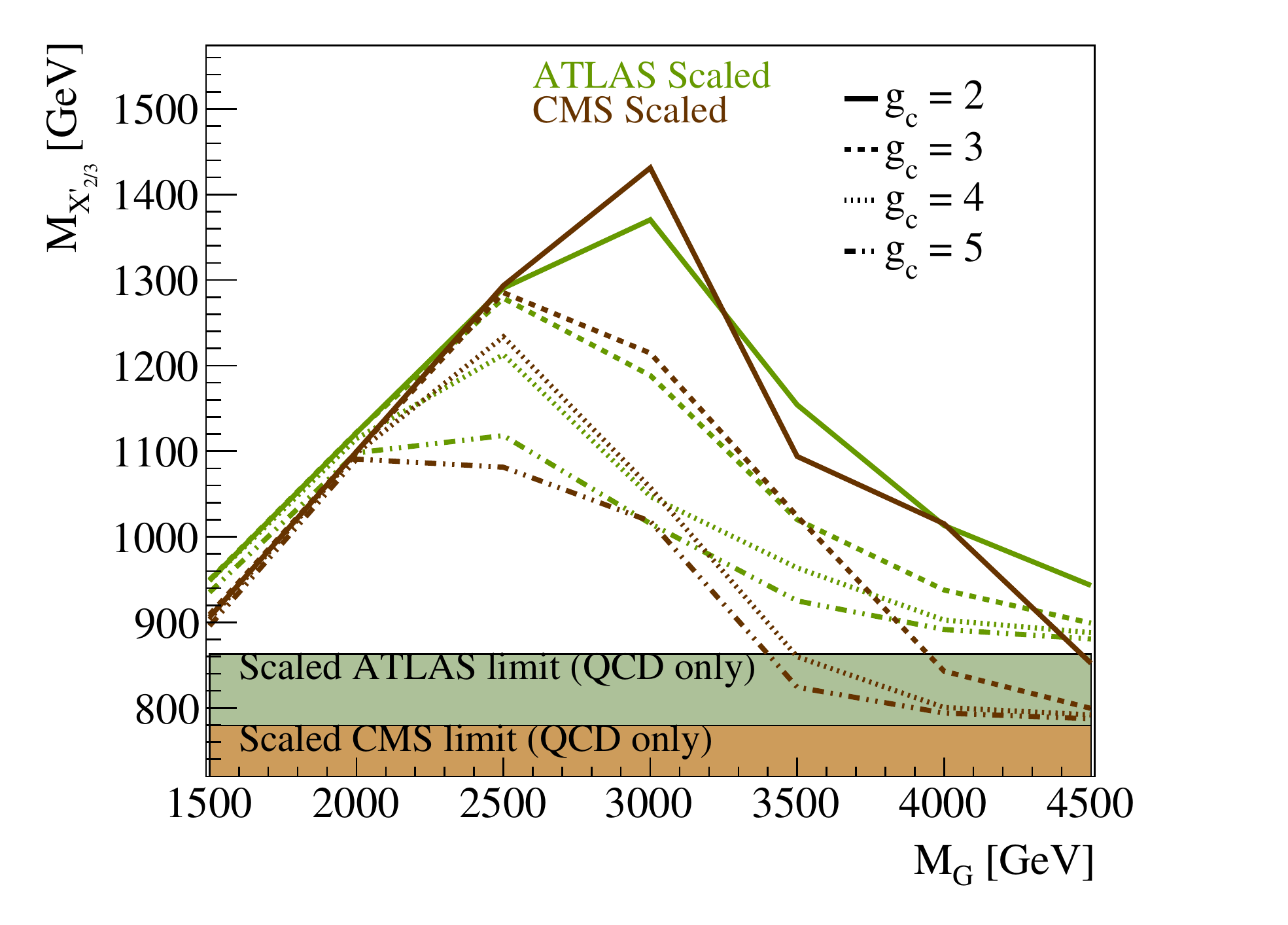}
\caption{}
\end{subfigure}
\begin{subfigure}[b]{0.49\textwidth}
\includegraphics[width=\textwidth]{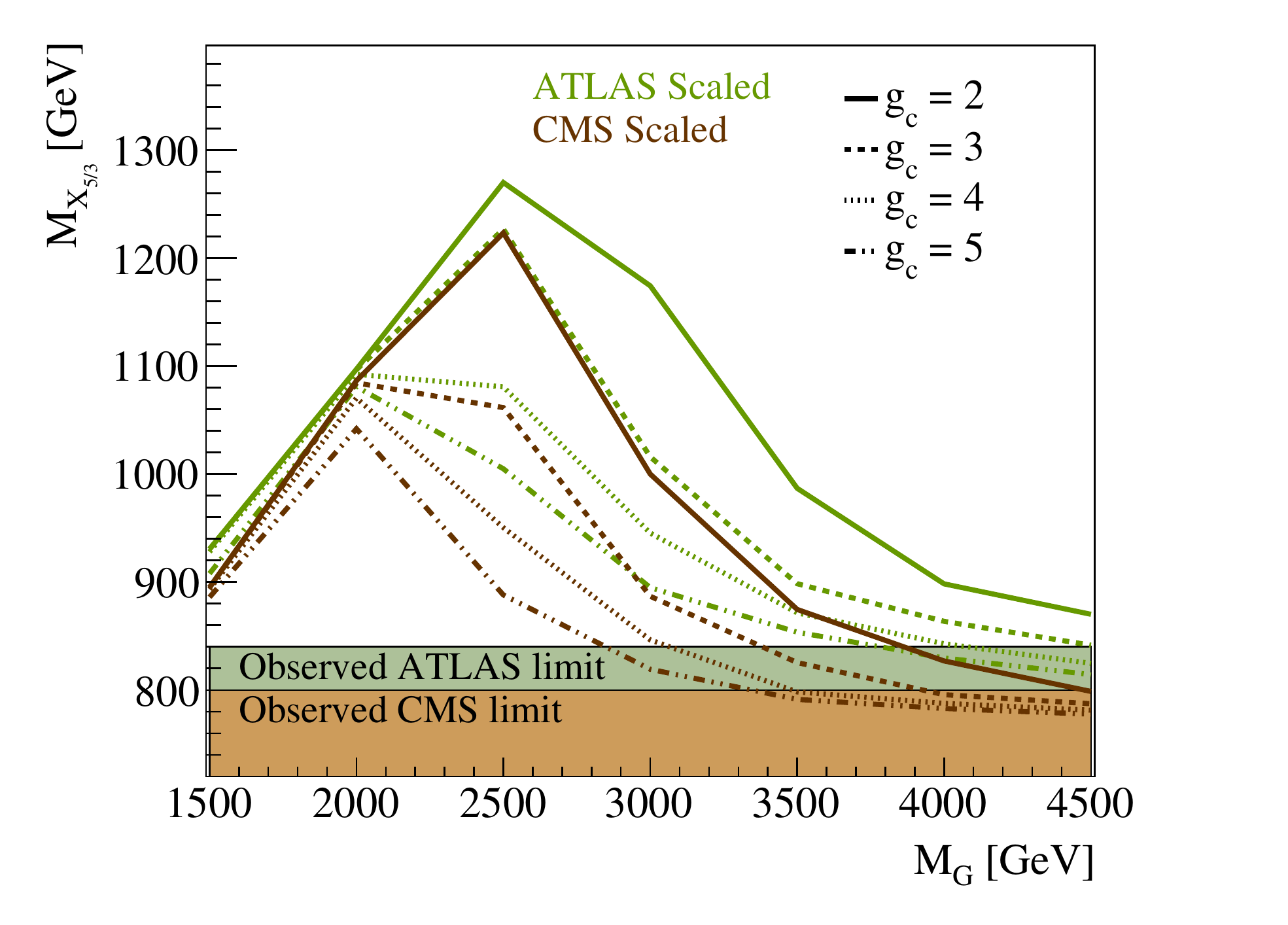}
\caption{}
\end{subfigure}
\end{center}
\caption{95\% C.L. lower  limits on the $M_G-M_{\Xpr}$ (left) and
$M_G-M_{X_{5/3}}$ (right) planes, derived from full run-1 published
data ($\sqrt{s}=8$~TeV), for different values of $g_c$. The filled orange and
green region represent the limits for CMS and ATLAS, respectively,
assuming QCD production of $\Xpr$ and $T^\prime$ (left) and $X_{5/3}$ (right).}
\label{fig:scaledht12}
\end{figure}

We have used the following experimental results to compute the current
limits on the VLQ present in out model.
For $B$ we have used the results published in~\cite{Aad:2015mba} (ATLAS)
and~\cite{CMS:2014cca} (CMS). 
The ATLAS
analysis considers a singlet model with different BR than in our
case. However, it is very sensitive to the $W$ decay and, as discussed,
we have
conservatively rescaled their result to our 100\% BR into $Wt$.
The CMS result is already published for a BR of 100\% to
the $W$ channel. Unfortunately, the mass of the $B$ quark in our model
is, for the range of parameters we have considered, typically heavier
than the masses these analyses are sensitive to and we cannot
therefore report bounds for all the $M_G$ values. 
We can nevertheless
provide the corresponding limits in some cases. For instance,
using the ATLAS results we can exclude values of $M_{B}$ ranging
from $M_{B}\simeq 750$~GeV for $M_{G} = 4$~TeV to 
$M_{B}\simeq 1$~TeV for $M_{G} = 2$~TeV. 
Using the CMS results we can excluded
values of $M_{B}$ ranging from $M_{B}\simeq 800$~GeV for 
$M_{G} = 3$~TeV to $M_{B}\simeq 950$~TeV for $M_{G} = 2$~TeV. 
These limits have been derived assuming $g_c=3$.
For the $\Xpr$ and $T'$ quarks we have used the results
in~\cite{Aad:2015kqa} (ATLAS) and~\cite{Chatrchyan:2013uxa} (CMS).  
The ATLAS results contain a combination of two analyses that target
decays of the VLQ into $H$ and $W$, respectively. Since charged current
decays are not present in our model we use the result quoted for the
doublet (same BR as in our model) 
only in the $H$ channel. The CMS analysis reports limits for a
2:1:1 BR pattern into $W$, $H$ and $Z$, respectively, and the analysis
used is very sensitive to the $Z$ channels. Thus, we conservatively
scale their results to out BR as discussed above. The results can be
found in the left panel of Fig.~\ref{fig:scaledht12} 
for different values of $g_c$. The corresponding value of
$M_{T^\prime}$ can be obtained from Fig.~\ref{fig:ht12compare}.
In the case of the $X_{5/3}$ the same ATLAS analysis used for the $B$
quark published results for the search of a vector-like quark with a
electric charge of $5/3$ assuming a BR of 100\% to
$Wt$. We have also used the CMS analysis~\cite{Chatrchyan:2013wfa} searching
for VLQ with an electric charge of $5/3$ in the same-sign
dilepton final state. They assume a 100\% BR into $Wt$ so we can
directly use their results, which can be found in the right panel of
Fig.~\ref{fig:scaledht12}.

\subsection{Early run-2 expectations\label{sec:early_run_2}}

Once we have computed the current bounds in the $M_G-M_Q$ plane with
$Q$ running over all the VLQ present in our model we would like to
compare them with the reach of a run-2 early data set. The goal is
two-fold, first we would like to see how much the limits can improve
with a few fb$^{-1}$ of integrated luminosity at $\sqrt{s}=13$ TeV
with respect to the current ones; second we would like to explore the
sensitivity to the presence of the heavy gluon at the initial stages
of a new run with higher energy, as opposed to the analyses we have
performed so far in which the limits are at the verge of the
kinematical reach.

\begin{figure}[t]
\begin{center}
\begin{subfigure}[b]{0.6\textwidth}
\includegraphics[width=\textwidth]{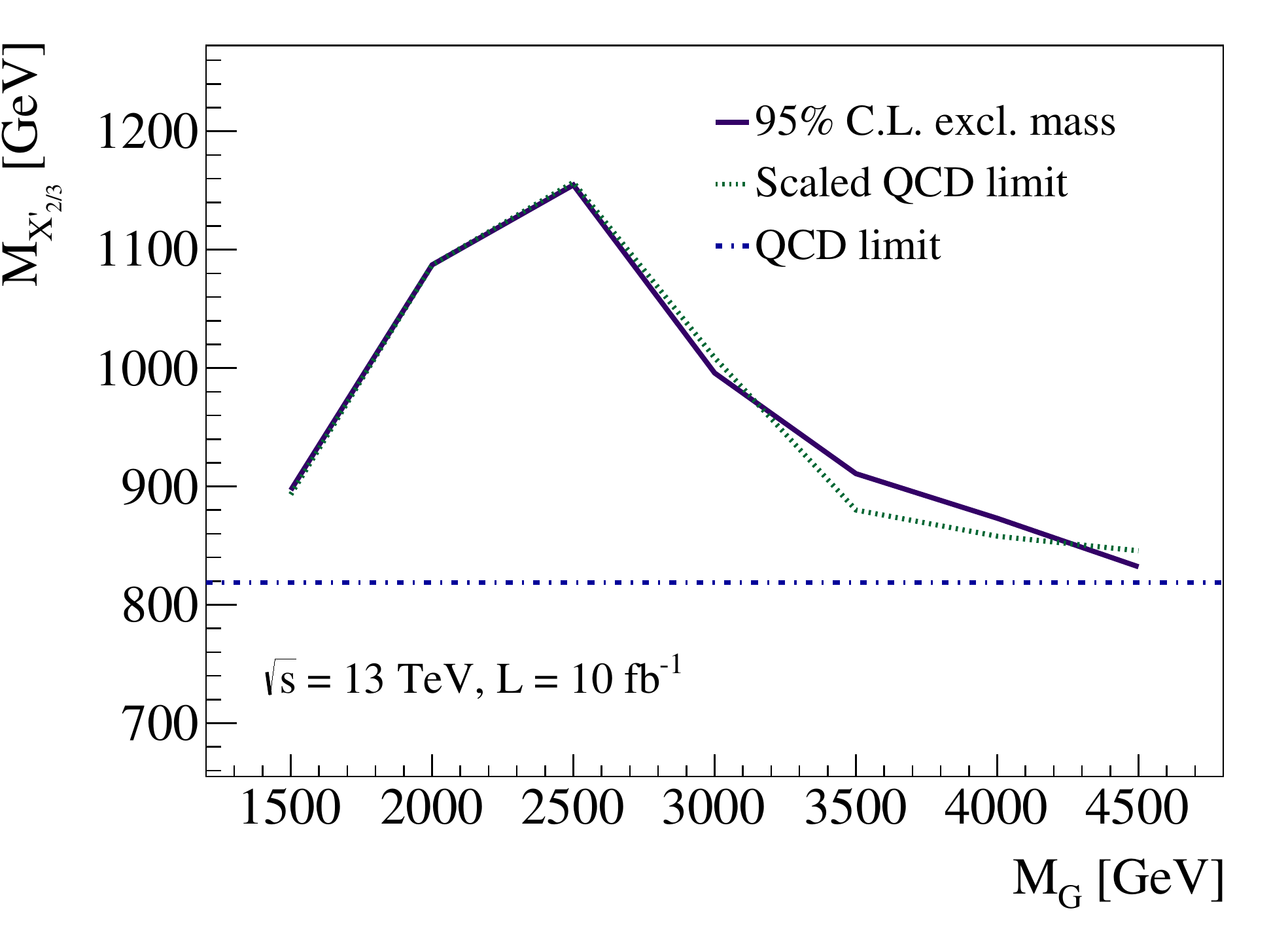}
\caption{}
\end{subfigure}
\end{center}
\caption{95\% C.L. lower limit on the $\Xpr$ mass as a function of
$M_G$ at $\sqrt{s}=13$~TeV and an integrated luminosity of
$10\text{~fb}^{-1}$. The green line shows the lower mass limit
assuming that the only effect is an increase in the cross section and
the purple line shows the lower mass limit when considering the
correct cross section upper limit for each pair of masses. The dashed
blue line shows the lower mass limit when only QCD production is
assumed.} 
\label{fig:13TeVmasslimits}
\end{figure}

We take as a benchmark analysis the multivariate one we presented in
Section~\ref{reconstruction} and estimate its reach in the
$M_G-M_{\Xpr}$ plane at
$\sqrt{s}=13$~TeV and an integrated luminosity of
$10\text{~fb}^{-1}$. We use the same selection and tools and compute
the limits again in two ways, either including QCD+HG from the start
or computing the upper limits in the cross section assuming QCD production
and rescaling the corresponding cross section.
Our results are summarized in Fig.~\ref{fig:13TeVmasslimits}, in which
we show the 95\% C.L. lower bounds in mass for the
$\Xpr$ quark as a function of the $G$ mass. Two points are worth
emphasizing. The first point is that the limits computed by a simple
rescaling of 
the cross section (using the results that assume QCD production as
would be published by the experimental collaborations) are perfectly
compatible with the full ones computed with the model at hand. Thus,
even when exploring regions of parameter space far from the
kinematical reach of the experiment, it seems like the experimental
limits on the pair production cross section of VLQ is quite
insensitive to the presence of the heavy gluon. It is therefore
perfectly legitimate to use them to put bounds on any model in which
heavy gluons contribute to the VLQ production.
A second important 
point is that there is, already with 10 fb$^{-1}$ of integrated
luminosity, a significant improvement in the limits with respect to
current ones. The early run-2 limits on the mass of the VLQ are in the
$\sim 820-1160$~GeV range (to be compared with the
equivalent $\sim 660-1000$ GeV at $\sqrt{s}=8$ TeV). A similar
improvement can be expected with respect to the complete current limits
presented in Fig.~\ref{fig:scaledht12}. 
Given these two points, it is clear that as soon as the experimental
collaborations publish their results on VLQ production during run-2,
they can be easily translated to bounds on the parameter space of
models with heavy vectors in the spectrum.

\section{Conclusions\label{conclusions}}

We have considered in this article the problem of reinterpretation of
experimental results in terms of models different from the ones used
by the experimental collaborations. Such reinterpretations are always
delicate since it is difficult to assess the impact of the new features
of the model in the final results. Assuming that the signatures of the
two models are reasonably similar, there are, broadly speaking, two
options. The first option is that the distribution for the kinematical
observables used in the analysis are very similar in both models. In
that case the experimental sensitivity (limit on the production cross
section as a function of the relevant scale) will be very similar for
both models and the limits can be computed by a simple rescaling of
the theoretical cross section. The alternative option is that some
observables show significant differences between the two models. The
reinterpretation of the limits in the new model become then much more
complicated, and the only possibility is to recast the
corresponding analysis using fast detector simulation tools.
This latter case is
not completely satisfactory, as it does not take full advantage of
the experimental publication and requires a certain degree of mastery
in the implementation of the model and the use of Monte Carlo and
detector simulation tools. On the other hand, it offers 
extra handles on how to more
efficiently search for the new model.

For the sake of concreteness we have considered pair production of new
VLQ in CHM. A common occurence in these models is
the presence of a heavy color octet vector boson, heavy gluon, that
can contribute to the pair production of the VLQ. We have performed a
detailed quantitative study of the effect that the heavy gluon
has on the distributions of different kinematical variables, both at
parton level and after detector simulation. The small differences
found at the parton level in variables that are relatively up in the
decay chain are almost completely washed out after detector
simulation. Thus, in practice, the limits on the VLQ and heavy gluon
masses that one obtains by using the full model and the ones that use
just QCD production (neglecting the heavy gluon) to compute the limit
on the production cross section are virtually identical. The latter
has the precious advantage that it uses directly the published
limits computed by the experimental collaborations and require only a minimal
amount of Monte Carlo simulations to compute the corresponding
theoretical cross sections at the parton level.

We have tested the validity of this approximation in two different
regimes of an experiment, one in which the limits are close to the
kinematical reach of the experiment and another in which a new energy
frontier is being explored for the first time. These correspond to the
full LHC run-1 data and the early run-2 data, respectively. In both
cases we have found that the experimental limits on the production
cross section are quite insensitive to the presence of the heavy gluon
and therefore the experimental results can easily be reinterpreted in
models with new VLQ and heavy gluons. This extends not only to current
analyses but also to more sophisticated analyses in which full
kinematical information, including the reconstruction of the VLQ, is
used. Presumably only a full reconstruction of the VLQ pair invariant
mass could give a significant sensitivity to the presence of the heavy
gluon. 

We have reached these conclusions by performing an exhaustive set of
comparisons of kinematical distributions and detailed
statistical analyses, which include multivariate
techniques. We have found that the main degradation of
the HG signal occurs at the reconstruction level, due in part to a
wrong assignment of physical objects by the reconstruction algorithms
in the presence of boosted topologies due to the overlap of the
considered jets. The use of boosted techniques
might partially compensate this effect but a detailed analysis, which
is beyond the scope of this study, would be needed to assess in a
quantitative way the discriminating power of a reconstruction method
using larger jets and jet substructure variables.
In any case, our modeling of the detector is based on fast
detector simulation and therefore our conclusions are limited by the
inherent accuracy of such a tool. In that sense it would be very
important that experimental collaborations consider the possibility of
comparing the experimental bounds on the VLQ pair production cross
section with and without a heavy gluon. 

We would also like to encourage the LHC experimental collaborations to
report their limits on the VLQ pair-production cross section as a
function of the VLQ BR.
They are available to the collaborations, as
they are used to compute the limits on the VLQ masses for arbitrary
BR. However they are not publicly available, thus limiting the
reinterpretation of their limits to other models, \emph{e.g.} with
or without heavy gluons.

\vspace*{0.3cm}
\noindent
\emph{Acknowledgements:} \\
We would like to thank G. Perez for useful discussions and motivation
at the initial stages of this work. 
JS is supported by MINECO, under grant
numbers FPA2010-17915 and FPA2013-47836-C3-2-P, 
by the European Commission through the contract
PITN-GA-2012-316704 (HIGGSTOOLS) and by Junta de Andaluc\'{\i}a grants
FQM 101 and FQM 6552. 
JS thanks the Pauli Center Visitor Program for financial support.
JPA and NFC are supported by FEDER, COMPETE-QREN and FCT, Portugal, 
through grant SFRH/BD/52002/2012 (JPA) and contract IF/00050/2013 (NFC).

\appendix

\section{Technical details of the model}\label{apend:model}

In this appendix we detail the different terms appearing in the 
Lagrangian Eq.(\ref{Lag:MCH45}).
The covariant derivatives read explicitly
\begin{eqnarray}
\mathrm{i}D_\mu q_L &=& 
\left(
\mathrm{i}\partial_\mu
+ g \frac{\sigma^i}{2} W^i_\mu + \frac{g^\prime}{6} B_\mu + g_e
G^e_\mu
\right)
q_L,
\nonumber \\
\mathrm{i}D_\mu t_R &=& 
\left(
\mathrm{i}\partial_\mu
+ \frac{2g^\prime}{3} B_\mu + g_c
G^c_\mu
\right)
t_R,
\nonumber \\
\mathrm{i}D_\mu \Psi &=& 
\left(
\mathrm{i}\partial_\mu
+ \frac{2g^\prime}{3} B_\mu + g_c
G^c_\mu
\right)
\Psi.
\end{eqnarray}
The remaining terms read
\begin{eqnarray}
\mathrm{i}\bar{\Psi}_R^i \cancel{d}_i t_R &=&
\frac{g}{\sqrt{2}} s_h [(\bar{X}_{5/3})_R \cancel{W}^+ 
-\bar{B}_R \cancel{W}^- ]t_R 
\nonumber \\
&&-\frac{g}{2c_W} s_h [\bar{T}_R + (\bar{X}_{2/3})_R]\cancel{Z} t_R
+\mathrm{i} [(\bar{X}_{2/3})_R - \bar{T}_R]
\frac{\cancel{\partial}\rho}{f} t_R,,
\\
\bar{\Psi}\left(\frac{2g^\prime}{3} \cancel{B} - \cancel{e}
\right)\Psi
&=&
\frac{g}{c_W}\left(-\frac{1}{2}+\frac{s_W^2}{3}\right) 
\bar{B} \cancel{Z} B
+\frac{g}{c_W}\left(\frac{1}{2}-\frac{5s_W^2}{3}\right) 
\bar{X}_{5/3} \cancel{Z} X_{5/3}
\nonumber \\
&+&
\frac{g}{c_W}\left(\frac{1}{2}c_h-\frac{2s_W^2}{3}\right) 
\bar{T} \cancel{Z} T
+
\frac{g}{c_W}\left(-\frac{1}{2}c_h-\frac{2s_W^2}{3}\right) 
\bar{X}_{2/3} \cancel{Z} X_{2/3}
\nonumber \\
&+& \frac{g}{\sqrt{2}}
\Big\{ 
\bar{B} \cancel{W}^- \left[ c_{h/2}^2 T + s_{h/2}^2 X_{2/3}\right]
+\bar{X}_{5/3} \cancel{W}^+ \left[ s_{h/2}^2 T + c_{h/2}^2 X_{2/3}\right]
+\mathrm{h.c.} \Big\}
\nonumber \\
&+& \mbox{ photon couplings},
\\
(\bar{Q}^5_L)^I U_{Ii} \Psi^i_R &=&
 \bar{b}_L B_R +  \bar{t}_L 
\Big[ c_{h/2}^2 T_R  + s_{h/2}^2 (X_{2/3})_R \Big], 
\\
 (\bar{Q}^5_L)^I U_{I5} t_R &=&
-\frac{1}{\sqrt{2}} s_h \bar{t}_L t_R,
\end{eqnarray}
where we have denoted
\begin{equation}
s_x \equiv \sin \frac{x}{f}, \quad
c_x \equiv \cos \frac{x}{f}, 
\end{equation}
except for $s_W$ and $c_W$, which are the sine and cosine of the
Weinberg angle. $\rho$ is the physical Higgs boson and $h$ reads, in
the unitary gauge
\begin{equation}
h \equiv \langle h \rangle + \rho,
\end{equation}
with 
\begin{equation}
f s_{\langle h \rangle} = v \approx 246\mbox{ GeV}.
\end{equation} 
These terms fix the electroweak couplings of the top and its partners
and also the top mass and mixing. In particular the latter two have
some implications on the degree of compositeness of the top quark and
therefore on its couplings to the heavy gluon.


\begin{thebibliography}{10}
\bibitem{Aad:2012tfa}
  G.~Aad {\it et al.} [ATLAS Collaboration],
  %``Observation of a new particle in the search for the Standard Model Higgs boson with the ATLAS detector at the LHC,''
  Phys.\ Lett.\ B {\bf 716} (2012) 1
  [arXiv:1207.7214 [hep-ex]].
  %%CITATION = ARXIV:1207.7214;%%
  %4730 citations counted in INSPIRE as of 16 juil. 2015
%\cite{Chatrchyan:2012xdj}

\bibitem{Chatrchyan:2012xdj}
  S.~Chatrchyan {\it et al.} [CMS Collaboration],
  %``Observation of a new boson at a mass of 125 GeV with the CMS experiment at the LHC,''
  Phys.\ Lett.\ B {\bf 716} (2012) 30
  [arXiv:1207.7235 [hep-ex]].
  %%CITATION = ARXIV:1207.7235;%%
  %4635 citations counted in INSPIRE as of 16 juil. 2015



  %\cite{Aad:2015kqa}

\bibitem{Aad:2015kqa} 
  G.~Aad {\it et al.}  [ATLAS Collaboration],
  %``Search for production of vector-like quark pairs and of four top quarks in the lepton-plus-jets final state in $pp$ collisions at $\sqrt{s}=8$ TeV with the ATLAS detector,''
  arXiv:1505.04306 [hep-ex].
  %%CITATION = ARXIV:1505.04306;%%
  %4 citations counted in INSPIRE as of 07 juil. 2015

  %\cite{Chatrchyan:2013uxa}

\bibitem{Chatrchyan:2013uxa} 
  S.~Chatrchyan {\it et al.}  [CMS Collaboration],
  %``Inclusive search for a vector-like T quark with charge $\frac{2}{3}$ in pp collisions at $\sqrt{s}$ = 8 TeV,''
  Phys.\ Lett.\ B {\bf 729}, 149 (2014)
  [arXiv:1311.7667 [hep-ex]].
  %%CITATION = ARXIV:1311.7667;%%
  %97 citations counted in INSPIRE as of 07 juil. 2015

%\cite{Alves:2011wf}

\bibitem{Alves:2011wf}
  D.~Alves {\it et al.} [LHC New Physics Working Group Collaboration],
  %``Simplified Models for LHC New Physics Searches,''
  J.\ Phys.\ G {\bf 39} (2012) 105005
  [arXiv:1105.2838 [hep-ph]].
  %%CITATION = ARXIV:1105.2838;%%
  %266 citations counted in INSPIRE as of 16 Jul 2015


%\cite{Kaplan:1983fs}

\bibitem{Kaplan:1983fs}
  D.~B.~Kaplan and H.~Georgi,
  %``SU(2) x U(1) Breaking by Vacuum Misalignment,''
  Phys.\ Lett.\ B {\bf 136} (1984) 183;
  %%CITATION = PHLTA,B136,183;%%
  %460 citations counted in INSPIRE as of 06 juil. 2015
%\cite{Kaplan:1983sm}
%
\bibitem{Kaplan:1991dc}
  D.~B.~Kaplan,
  %``Flavor at SSC energies: A New mechanism for dynamically generated fermion masses,''
  Nucl.\ Phys.\ B {\bf 365} (1991) 259;
  %%CITATION = NUPHA,B365,259;%%
  %181 citations counted in INSPIRE as of 16 juil. 2015
%\cite{Contino:2006nn}
%
\bibitem{Matsedonskyi:2012ym}
  O.~Matsedonskyi, G.~Panico and A.~Wulzer,
  %``Light Top Partners for a Light Composite Higgs,''
  JHEP {\bf 1301} (2013) 164
  [arXiv:1204.6333 [hep-ph]];
  %%CITATION = ARXIV:1204.6333;%%
  %101 citations counted in INSPIRE as of 16 juil. 2015
%\cite{Redi:2012ha}
%
\bibitem{Agashe:2003zs}
  K.~Agashe, A.~Delgado, M.~J.~May and R.~Sundrum,
  %``RS1, custodial isospin and precision tests,''
  JHEP {\bf 0308} (2003) 050
  [hep-ph/0308036];
  %%CITATION = HEP-PH/0308036;%%
  %520 citations counted in INSPIRE as of 06 Jul 2015
%\cite{Carena:2006bn}
%
\bibitem{Chala:2014mma}
  M.~Chala, J.~Juknevich, G.~Perez and J.~Santiago,
  %``The Elusive Gluon,''
  JHEP {\bf 1501} (2015) 092
  [arXiv:1411.1771 [hep-ph]].
  %%CITATION = ARXIV:1411.1771;%%
  %5 citations counted in INSPIRE as of 24 Apr 2015

%\cite{Barcelo:2011wu}

\bibitem{Barcelo:2011wu}
  R.~Barcelo, A.~Carmona, M.~Chala, M.~Masip and J.~Santiago,
  %``Single Vectorlike Quark Production at the LHC,''
  Nucl.\ Phys.\ B {\bf 857} (2012) 172
  [arXiv:1110.5914 [hep-ph]].
  %%CITATION = ARXIV:1110.5914;%%
  %35 citations counted in INSPIRE as of 16 juil. 2015

%\cite{Bini:2011zb}

\bibitem{Bini:2011zb}
  C.~Bini, R.~Contino and N.~Vignaroli,
  %``Heavy-light decay topologies as a new strategy to discover a heavy gluon,''
  JHEP {\bf 1201} (2012) 157
  [arXiv:1110.6058 [hep-ph]];
  %%CITATION = ARXIV:1110.6058;%%
  %36 citations counted in INSPIRE as of 16 Jul 2015
%\cite{Carmona:2012jk}
%
\bibitem{Azatov:2015xqa}
  A.~Azatov, D.~Chowdhury, D.~Ghosh and T.~S.~Ray,
  %``Same sign di-lepton candles of the composite gluons,''
  arXiv:1505.01506 [hep-ph].
  %%CITATION = ARXIV:1505.01506;%%
  %2 citations counted in INSPIRE as of 08 Jul 2015


%\cite{Vignaroli:2015ama}
\bibitem{Vignaroli:2015ama}
  N.~Vignaroli,
  %``$Z$-peaked excess from heavy gluon decays to vectorlike quarks,''
  Phys.\ Rev.\ D {\bf 91} (2015) 11,  115009
  [arXiv:1504.01768 [hep-ph]].
  %%CITATION = ARXIV:1504.01768;%%
  %9 citations counted in INSPIRE as of 24 juil. 2015

%\cite{Agashe:2004rs}

\bibitem{Agashe:2004rs}
  K.~Agashe, R.~Contino and A.~Pomarol,
  %``The Minimal composite Higgs model,''
  Nucl.\ Phys.\ B {\bf 719} (2005) 165
  [hep-ph/0412089];
  %%CITATION = HEP-PH/0412089;%%
  %692 citations counted in INSPIRE as of 24 Apr 2015
%\cite{Contino:2006qr}
%
\bibitem{DeSimone:2012fs}
  A.~De Simone, O.~Matsedonskyi, R.~Rattazzi and A.~Wulzer,
  %``A First Top Partner Hunter's Guide,''
  JHEP {\bf 1304} (2013) 004
  [arXiv:1211.5663 [hep-ph]].
  %%CITATION = ARXIV:1211.5663;%%
  %95 citations counted in INSPIRE as of 24 Apr 2015

%\cite{Barcelo:2011vk}

\bibitem{Barcelo:2011vk}
  R.~Barcelo, A.~Carmona, M.~Masip and J.~Santiago,
  %``Stealth gluons at hadron colliders,''
  Phys.\ Lett.\ B {\bf 707} (2012) 88
  [arXiv:1106.4054 [hep-ph]].
  %%CITATION = ARXIV:1106.4054;%%
  %58 citations counted in INSPIRE as of 24 Apr 2015

  

%\cite{Greco:2014aza}

\bibitem{Greco:2014aza}
  D.~Greco and D.~Liu,
  %``Hunting composite vector resonances at the LHC: naturalness facing data,''
  JHEP {\bf 1412} (2014) 126
  [arXiv:1410.2883 [hep-ph]].
  %%CITATION = ARXIV:1410.2883;%%
  %9 citations counted in INSPIRE as of 08 juil. 2015


%\cite{Carmona:2015xaa}

\bibitem{Carmona:2015xaa}
  A.~Carmona, A.~Delgado, M.~Quiros and J.~Santiago,
  %``Diboson resonant production in non-custodial composite Higgs models,''
  arXiv:1507.01914 [hep-ph].
  %%CITATION = ARXIV:1507.01914;%%

  %\cite{Aad:2014efa}

\bibitem{Aad:2014efa} 
  G.~Aad {\it et al.}  [ATLAS Collaboration],
  %``Search for pair and single production of new heavy quarks that decay to a $Z$ boson and a third-generation quark in $pp$ collisions at $\sqrt{s}=8$ TeV with the ATLAS detector,''
  JHEP {\bf 1411}, 104 (2014)
  [arXiv:1409.5500 [hep-ex]].
  %%CITATION = ARXIV:1409.5500;%%
  %20 citations counted in INSPIRE as of 13 May 2015
  
%\cite{Degrande:2011ua}

\bibitem{Degrande:2011ua}
  C.~Degrande, C.~Duhr, B.~Fuks, D.~Grellscheid, O.~Mattelaer and T.~Reiter,
  %``UFO - The Universal FeynRules Output,''
  Comput.\ Phys.\ Commun.\  {\bf 183} (2012) 1201
  [arXiv:1108.2040 [hep-ph]].
  %%CITATION = ARXIV:1108.2040;%%
  %210 citations counted in INSPIRE as of 16 juil. 2015


%\cite{Alloul:2013bka}

\bibitem{Alloul:2013bka}
  A.~Alloul, N.~D.~Christensen, C.~Degrande, C.~Duhr and B.~Fuks,
  %``FeynRules  2.0 - A complete toolbox for tree-level phenomenology,''
  Comput.\ Phys.\ Commun.\  {\bf 185} (2014) 2250
  [arXiv:1310.1921 [hep-ph]].
  %%CITATION = ARXIV:1310.1921;%%
  %217 citations counted in INSPIRE as of 16 juil. 2015


%\cite{Alwall:2014hca}

\bibitem{Alwall:2014hca}
  J.~Alwall {\it et al.},
  %``The automated computation of tree-level and next-to-leading order differential cross sections, and their matching to parton shower simulations,''
  JHEP {\bf 1407} (2014) 079
  [arXiv:1405.0301 [hep-ph]].
  %%CITATION = ARXIV:1405.0301;%%
  %397 citations counted in INSPIRE as of 16 Jul 2015

%\cite{Sjostrand:2006za}

\bibitem{Sjostrand:2006za}
  T.~Sjostrand, S.~Mrenna and P.~Z.~Skands,
  %``PYTHIA 6.4 Physics and Manual,''
  JHEP {\bf 0605} (2006) 026
  [hep-ph/0603175].
  %%CITATION = HEP-PH/0603175;%%
  %6499 citations counted in INSPIRE as of 16 juil. 2015

  %\cite{deFavereau:2013fsa}

\bibitem{deFavereau:2013fsa} 
  J.~de Favereau {\it et al.}  [DELPHES 3 Collaboration],
  %``DELPHES 3, A modular framework for fast simulation of a generic collider experiment,''
  JHEP {\bf 1402}, 057 (2014)
  [arXiv:1307.6346 [hep-ex]].
  %%CITATION = ARXIV:1307.6346;%%
  %230 citations counted in INSPIRE as of 13 May 2015
  

 %\cite{Cacciari:2011ma}

\bibitem{Cacciari:2011ma}
  M.~Cacciari, G.~P.~Salam and G.~Soyez,
  %``FastJet User Manual,''
  Eur.\ Phys.\ J.\ C {\bf 72} (2012) 1896
  [arXiv:1111.6097 [hep-ph]].
  %%CITATION = ARXIV:1111.6097;%%
  %915 citations counted in INSPIRE as of 16 Jul 2015 
  

%\cite{Artoisenet:2012st}

\bibitem{Artoisenet:2012st}
  P.~Artoisenet, R.~Frederix, O.~Mattelaer and R.~Rietkerk,
  %``Automatic spin-entangled decays of heavy resonances in Monte Carlo simulations,''
  JHEP {\bf 1303} (2013) 015
  [arXiv:1212.3460 [hep-ph]].
  %%CITATION = ARXIV:1212.3460;%%
  %43 citations counted in INSPIRE as of 16 juil. 2015

%\cite{Hocker:2007ht}

\bibitem{Hocker:2007ht}
  A.~Hocker {\it et al.},
  %``TMVA - Toolkit for Multivariate Data Analysis,''
  PoS ACAT (2007) 040
  [physics/0703039].
  %%CITATION = PHYSICS/0703039;%%
  %523 citations counted in INSPIRE as of 16 juil. 2015

%\cite{Read:2002hq}

\bibitem{Read:2002hq}
  A.~L.~Read,
  %``Presentation of search results: The CL(s) technique,''
  J.\ Phys.\ G {\bf 28} (2002) 2693.
  %%CITATION = JPAGA,G28,2693;%%
  %1152 citations counted in INSPIRE as of 16 juil. 2015


%\cite{AguilarSaavedra:2009es}

\bibitem{AguilarSaavedra:2009es}
  J.~A.~Aguilar-Saavedra,
  %``Identifying top partners at LHC,''
  JHEP {\bf 0911} (2009) 030
  [arXiv:0907.3155 [hep-ph]];
  %%CITATION = ARXIV:0907.3155;%%
  %134 citations counted in INSPIRE as of 16 juil. 2015
%\cite{Matsedonskyi:2014mna}
%
\bibitem{Aad:2015mba} 
  G.~Aad {\it et al.}  [ATLAS Collaboration],
  %``Search for vectorlike $B$ quarks in events with one isolated lepton, missing transverse momentum and jets at $\sqrt{s}=$ 8 TeV with the ATLAS detector,''
  Phys.\ Rev.\ D {\bf 91}, no. 11, 112011 (2015)
  [arXiv:1503.05425 [hep-ex]].
  %%CITATION = ARXIV:1503.05425;%%
  %6 citations counted in INSPIRE as of 07 juil. 2015
  
  
\bibitem{CMS:2014cca} 
  CMS Collaboration,
  %``Search for a vector-like bottom quark partner in same sign di-lepton final states,''
  CMS-PAS-B2G-12-020.
  %%CITATION = CMS-PAS-B2G-12-020;%%
  %2 citations counted in INSPIRE as of 16 juil. 2015

\bibitem{Chatrchyan:2013wfa} 
  S.~Chatrchyan {\it et al.}  [CMS Collaboration],
  %``Search for top-quark partners with charge 5/3 in the same-sign dilepton final state,''
  Phys.\ Rev.\ Lett.\  {\bf 112}, no. 17, 171801 (2014)
  [arXiv:1312.2391 [hep-ex]].
  %%CITATION = ARXIV:1312.2391;%%
  %45 citations counted in INSPIRE as of 08 juil. 2015
  
  %\cite{CMS:2014cca}

\end{thebibliography}
\end{document}